\documentclass[11pt,a4paper]{article}
\pdfoutput=1
\usepackage{jheppub}
\usepackage{slashed}
\usepackage{multirow}
\usepackage{color}
\usepackage{mathtools}

\newcommand{\slv}{\raise.15ex\hbox{$/$}\kern-.53em\hbox{$v$}}
\newcommand{\slF}{\raise.15ex\hbox{$/$}\kern-.53em\hbox{$F$}}
\newcommand{\slL}{\raise.15ex\hbox{$/$}\kern-.53em\hbox{$L$}}
\newcommand{\slP}{\raise.15ex\hbox{$/$}\kern-.53em\hbox{$P$}}
\newcommand{\slp}{\raise.15ex\hbox{$/$}\kern-.53em\hbox{$p$}}
\newcommand{\slq}{\raise.15ex\hbox{$/$}\kern-.53em\hbox{$q$}}
\newcommand{\slR}{\raise.15ex\hbox{$/$}\kern-.53em\hbox{$R$}}
\newcommand{\slQ}{\raise.15ex\hbox{$/$}\kern-.53em\hbox{$Q$}}
\newcommand{\slK}{\raise.15ex\hbox{$/$}\kern-.53em\hbox{$K$}}
\newcommand{\slk}{\raise.15ex\hbox{$/$}\kern-.53em\hbox{$k$}}
\newcommand{\slD}{\raise.15ex\hbox{$/$}\kern-.53em\hbox{$D$}}
\newcommand{\slC}{\raise.15ex\hbox{$/$}\kern-.53em\hbox{$C$}}
\newcommand{\slA}{\raise.15ex\hbox{$/$}\kern-.53em\hbox{$A$}}
\newcommand{\slSigma}{\raise.15ex\hbox{$/$}\kern-.53em\hbox{$\Sigma$}}
\newcommand{\slpartial}{\raise.15ex\hbox{$/$}\kern-.53em\hbox{$\partial$}}
\newcommand{\slcalP}{\raise.15ex\hbox{$/$}\kern-.63em\hbox{$\cal P$}}

\def\bs{\boldsymbol}

\def\bdel{\boldsymbol \partial}

\def\p{{\boldsymbol p}}

\def\q{{\boldsymbol q}}

\def\k{{\boldsymbol k}}

\def\n{{\boldsymbol n}}

\def\y{{\boldsymbol y}}

\def\r{{\boldsymbol r}}
\def\z{{\boldsymbol z}}

\def\bkappa{{\boldsymbol \kappa}}
\def\bbkappa{\bar{\boldsymbol \kappa}}

\def\K{\hat{\boldsymbol k}_{1}}

\def\qqb{{q\bar q}}

\def\inin{{in-in}}
\def\inout{{in-out}}
\def\outout{{out-out}}
\def\pip{\delta {\bs n}}
\def\thmed{\theta_\text{max}}
\def\tform{t_\text{f}}
\def\tdip{t_\text{d}}
\def\thform{\theta_\text{f}}

\def\kform{k_\text{f}}
\def\thcoh{\theta_\text{coh}}
\def\kcoh{k_\text{coh}}
\def\thcrit{\theta_\text{c}}
\def\Qdip{r_\perp^{-1}}
\def\Qmed{Q_s}
\def\Qhard{Q_\text{hard}}

\def\Rq{\mathcal{R}_{q}}
\def\Rqb{\mathcal{R}_{\bar q}}
\def\Rcoh{\mathcal{R}_\text{sing}}
\def\J{\mathcal{J}}

\def\K{\mathcal{K}}
\def\P{\mathcal{P}}

\newcommand{\beq}{\begin{eqnarray}}
\newcommand{\eeq}{\end{eqnarray}}
\newcommand{\be}{\begin{eqnarray*}}
\newcommand{\ee}{\end{eqnarray*}}

\newcommand{\nn}{\nonumber\\ }

\renewcommand{\tabcolsep}{0.5cm}

\title{The radiation pattern of a QCD antenna \\ in a dense medium}

\author[a]{Yacine Mehtar-Tani,}

\author[b,c]{Carlos A. Salgado}

\author[d]{and Konrad Tywoniuk}

\affiliation[a]{Institut de Physique Th\'eorique, CEA Saclay, 
F-91191 Gif-sur-Yvette, France}

\affiliation[b]{
Departamento de F\'isica de Part\'iculas and IGFAE,
Universidade de Santiago de Compostela, \\
E-15 782 Santiago de Compostela, Galicia-Spain}

\affiliation[c]{
Physics Department, Theory Unit, CERN, CH-1211 Gen\`eve 23, Switzerland
}

\affiliation[d]{
Department of Astronomy and Theoretical Physics,
Lund University,
S\"olvegatan 14A, 
SE-22 362 Lund, Sweden}

\emailAdd{yacine.mehtar-tani@cea.fr}
\emailAdd{carlos.salgado@usc.es}
\emailAdd{konrad.tywoniuk@thep.lu.se}

\abstract{
We calculate the radiation spectrum off a $q\bar q$ pair of a fixed opening angle $\theta_\qqb$ traversing a medium of length $L$. Multiple interactions with the medium are handled in the harmonic oscillator approximation, valid for soft gluon emissions. We discuss the time-scales relevant to the decoherence of correlated partons traversing the medium and demonstrate how this relates to the hard scale that govern medium-induced radiation. For large angle radiation, the hard scale is given by $\Qhard = \max\left(\Qdip, \Qmed \right)$, where $r_\perp = \theta_\qqb L$ is the probed transverse  size and $\Qmed$ is the maximal transverse momentum accumulated by the emitted gluon in the medium. These situations define in turn two distinct regimes, which we call ``dipole" and ``decoherence" regimes, respectively, which are discussed in detail. A feature common to both cases is that coherence of the radiation is restored at large transverse momenta, $k_\perp > \Qhard$.
}

\preprint{CERN-PH-TH/2012-144, LU-TP 11-24}

\begin{document}
\maketitle

\section{Introduction}
\label{sec:intro}

One has observed striking effects of energy loss of leading partons in heavy-ion collisions at RHIC and LHC, see e.g. \cite{d'Enterria:2009am,Majumder:2010qh} and references therein. These observations established the phenomenon known as `jet quenching,' namely the interaction of a hard parton with the dense medium created in the aftermath of the collision, as one of the key discoveries of the high-energy heavy-ion program \cite{Back:2004je,Arsene:2004fa,Adcox:2004mh,Adams:2005dq}. Complementary, more differential observables such as multi-particle correlations and full jet reconstruction in heavy-ion events probe the structure of the energy and particle distributions of jets traversing the quark-gluon plasma \cite{Putschke:2008wn,Salur:2008hs,Lai:2009zq,Aamodt:2010jd,Aad:2010bu,Chatrchyan:2011sx,Chatrchyan:2012ni} and provide strong constraints on the theoretical modeling of multi-gluon emissions in the medium. 

Usually, these effects are interpreted within the framework of induced radiative processes in the presence of a deconfined medium \cite{Gyulassy:1993hr}.\footnote{Elastic energy loss becomes important for the less energetic constituents of a jet, but is beyond the scope of the present work.} Until recently one has studied these theoretically on the level of the single-gluon emission spectrum which accounts for multiple scattering of both the projectile and emitted gluon with the medium. This approach was pioneered by the works of Baier-Dokshitzer-Mueller-Peigne-Schiff and Zakharov (BDMPS-Z) \cite{Baier:1996kr,Baier:1996sk,Baier:1998kq,Zakharov:1996fv,Zakharov:1997uu}. Further developments were presented in \cite{Wiedemann:1999fq,Wiedemann:2000za,Wiedemann:2000tf,Gyulassy:2000fs,Gyulassy:2000er,Arnold:2001ba,Arnold:2001ms,Arnold:2002ja}, see also \cite{MehtarTani:2006xq}. For other approaches, see \cite{Majumder:2010qh} and references therein. These studies are QCD extensions of the analogous QED problem of induced radiation, considered long ago by Landau, Pomeranchuk and Migdal \cite{Landau:1953um,Migdal:1956tc}. The medium-induced single-inclusive spectrum off a single charge, known henceforth as the BDMPS-Z spectrum, serves also as a building block for treating multi-gluon emissions, see, e.g., \cite{Baier:2001yt,Salgado:2003gb}. These {\it ad hoc} extensions can therefore only account for uncorrelated emissions (except for trivial correlations such as energy-momentum conservation) and serve as working models for phenomenological applications and Monte-Carlo implementations, e.g., \cite{Armesto:2009fj,Lokhtin:2008xi,Schenke:2009gb,Zapp:2011ya}.

A pressing question is whether this picture is valid and adequate for comparisons with data. On one hand, interferences are crucial in order to establish the relevant order variables for jet showering in vacuum at leading logarithmic accuracy. On the other hand, the BDMPS-Z spectrum is infrared and collinear safe and one does not expect any logarithmic enhancement of the spectrum. It is instead dominated by the characteristic transverse momentum accumulated via interactions with the medium, see section~\ref{sec:independent-analysis} for details,  

In our recent papers \cite{MehtarTani:2010ma, MehtarTani:2011tz, MehtarTani:2011gf}, see also \cite{CasalderreySolana:2011rz}, we have tried to address these questions from first principles and have pointed out interesting aspects arising when considering interference processes in the medium. In particular, one recovers an infrared divergent vacuum-like component which is responsible for the onset of decoherence of the vacuum radiation in the soft sector \cite{MehtarTani:2010ma}. We also showed how the onset of decoherence in a dense medium result in independent radiation off all particles \cite{MehtarTani:2011tz}. This can be interpreted as a  screening effect which also implies the `loss of memory' of color connections between parents and offspring in the radiative process.

The extension of this phenomenon to arbitrary gluon energies follows a general picture established in \cite{MehtarTani:2011gf}, see also \cite{Armesto:2011ir} for a treatment of the case of massive quarks. One can identify two distinct regimes characterizing the induced radiation which involve different physical processes but, strikingly, share several common features. This unifying idea follows from considering the relevant scales of the problem in the presence of a medium. In terms of characteristic transverse distances, one one hand there is the antenna size, denoted by $r_\perp$, and, on the other hand, the transverse color correlation length of the medium, given by the inverse of the characteristic medium scale $\Qmed^{-1}$.\footnote{$\Qmed$ is related to the maximum transverse momentum that a induced particle can accumulate traversing the total medium length. See sec.~\ref{sec:independent-analysis} for details.} Let us first consider the situation, when $r_\perp < \Qmed^{-1}$, i.e., the limit of small antenna sizes. In this case the antenna interacts as a coherent ensemble with the medium, since the medium cannot resolve its inner structure. In fact, in this case one can instead say that the medium is probed by the antenna. It follows that the medium interactions only can stimulate coherent radiation. This induces a vacuum-like radiation component at angles larger than the opening angle of the pair which modifies the angular ordering condition. This characterizes the so-called ``dipole" regime. In the opposite case, $r_\perp > \Qmed^{-1}$, the medium probes the inner structure of the antenna. We have dubbed this the ``decoherence" regime. Interferences are strongly suppressed and independent radiation is induces off each of the constituents. In both cases, the spectrum is governed by the hardest of these two momentum scales, $\Qhard = \max\left(\Qdip,\Qmed \right)$, which specifies the maximal transverse momentum of produced gluons. Most importantly, vacuum coherence is restored for $k_\perp > \Qhard$.

The analysis in \cite{MehtarTani:2011gf} was limited to consider only one scattering center inside the medium and is therefore strictly valid for a relatively dilute medium. The size of the medium, $L$, becomes in this case rather an indication of the longitudinal position of the scattering center. Besides, for this calculation the ``decoherence" regime does not involve the complete disappearance of interferences in the vacuum. In this framework, on the other hand, it is feasible to account analytically for effects of finite mean free path in the medium. In the present work we extend our previous efforts by considering an arbitrary number of re-scatterings with the medium. This allows us to analyze the corresponding interference effects in opaque media. But in order to gain some analytical insight we will employ an approximation which is only valid in the limit of vanishing mean free path --- the `harmonic oscillator' approximation --- strictly valid only for soft gluon production. Our present analysis will also help clarify the meaning of the relevant time-scales, such as the formation and emission times, in a continuous medium of a certain extension.

The outline of the paper is the following. In section~\ref{sec:formal-spectrum}, we give a brief outlook of the antenna radiation spectrum in vacuum and medium, discussing the main features of the independent components, see eq.~(\ref{eq:independentspec-main}), and the interferences, see eq.~(\ref{eq:interferencespec-main}). A general discussion about the physics of decoherence of QCD radiation is given in section~\ref{sec:discussion}, leading up to a more detailed discussion of the features of the spectrum in the following sections. In particular, the physics governing the two distinct regimes, namely the ``dipole" and ``decoherence" regimes, relevant for our setup, is explained and the main scales of the problem identified in secs.~\ref{sec:discussion-dipole} and \ref{sec:discussion-decoherence}, respectively. We introduce the harmonic oscillator approximation in section~\ref{sec:ho-approximation}, which allow us to simplify our expressions considerably. Then, as a first advance toward understanding interference effects in the medium we analyze the independent component of the spectrum which is less complicated in section~\ref{sec:independent-analysis}. In particular, we derive a novel formulation of the independent component, given by eq.~(\ref{eq:r-in-in-final-leading}). The insights gained in finding this leading behavior allow us to devise a simplified procedure, discussed in detail in section~\ref{sec:indep-revisited}, which aid considerably in simplifying the complicated structure of the interference spectrum. We analyze this in detail in section~\ref{sec:interferences} --- the ``dipole" regime is discussed in sec.~\ref{sec:interferences-dipole} and the ``decoherence" regime in sec.~\ref{sec:interferences-dense} --- and, throughout, substantiate the general picture outlined in sec.~\ref{sec:discussion}. The numerical results, serving to illustrate the aforementioned features, are presented in section~\ref{sec:numerics}. We supplement our discussion by including the possible non-zero total color charge of the antenna in sec.~\ref{sec:octet} and consider briefly the implications of our findings for a parton shower in the medium. Finally, we conclude and give a brief outlook in section~\ref{sec:conclusions}.

\section{The antenna radiation spectrum}
\label{sec:formal-spectrum}

\begin{figure}
\centering
\includegraphics[width=0.3\textwidth]{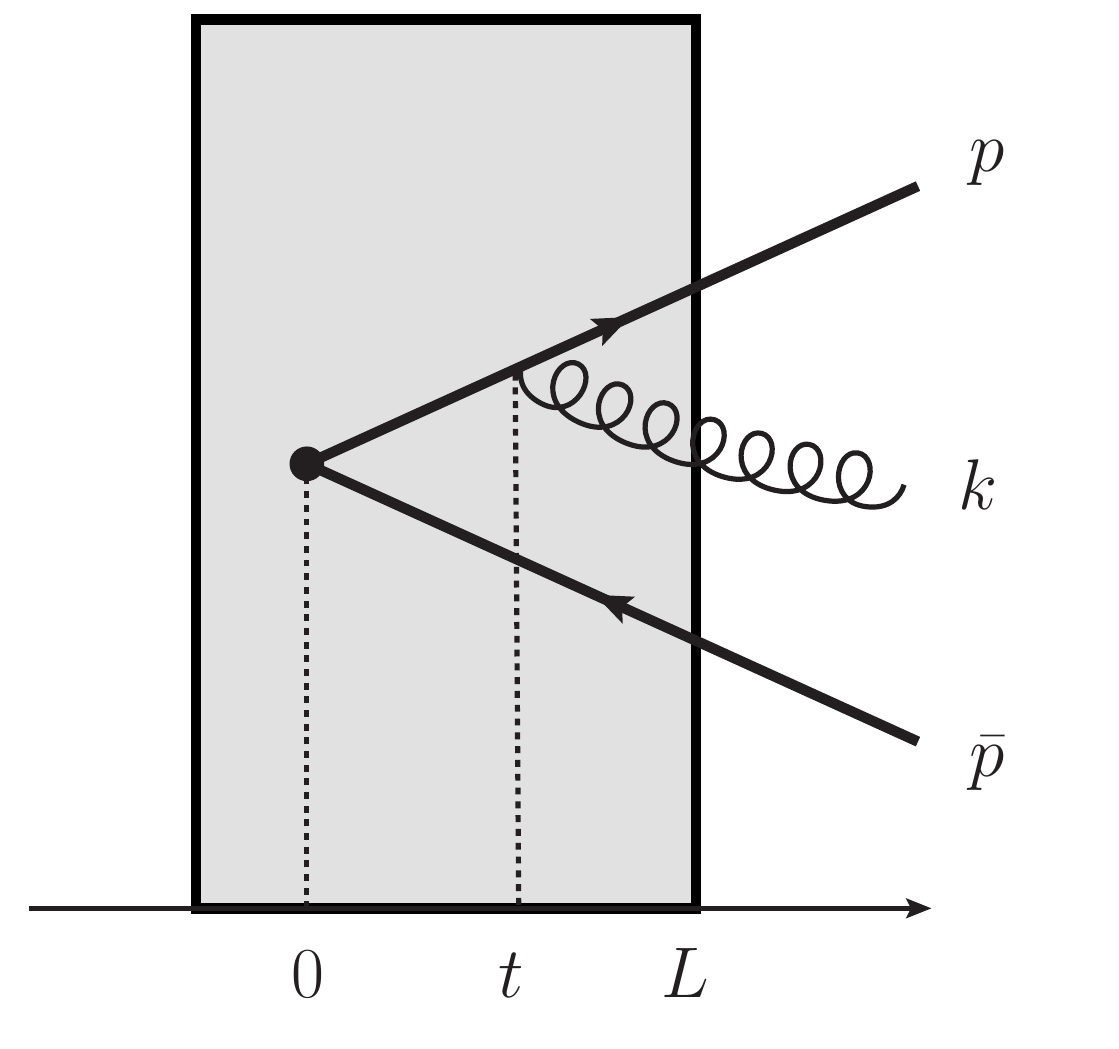}
\caption{The kinematics of the gluon emission off the $\qqb$ antenna.}
\label{fig:AntennaKinematics}
\end{figure}
Following closely the setup already studied in our previous works \cite{MehtarTani:2010ma, MehtarTani:2011tz, MehtarTani:2011gf}, we will analyze the radiation spectrum of a $\qqb$-pair, with a given opening angle $\theta_\qqb$, traversing an extended medium of free, uncorrelated color charges. The antenna originates either from the decay of a virtual photon or gluon. In the former case, the $\qqb$-pair is created in a color singlet state while in the latter as a color octet. Assuming, for the purpose of our discussion, that the quark and antiquark are very energetic implies a large virtuality of the initial projectile such that we can neglect emissions prior to the $\qqb$-splitting, in line with the leading logarithmic approximation (LLA), see, e.g., \cite{Dokshitzer:1991wu}. The spectrum of emitted gluons with energy $\omega$\footnote{The light-cone variables, $x^\pm \equiv (x^0 \pm x^3)/\sqrt{2}$, related to the longitudinal propagation and energy are simply denoted as $t \equiv x^+$ and $\omega \equiv k^+$, respectively, to ease the notation. Bold-face characters, e.g. ${\bs k}$, will denote two-dimensional transverse vectors.} and transverse momentum $\k$ reads in the general, octet case, 
\beq
\label{eq:spec-general}
\omega\frac{dN}{d^3k}=\frac{\alpha_s}{(2\pi)^2\,\omega^2}\ \big( C_F \Rcoh+C_A \mathcal{J} \big) \,,
\eeq
where $\Rcoh = \mathcal{R}_q + \mathcal{R}_{\bar q} - 2 \mathcal{J}$ is the color singlet component. The spectrum consists of independent radiation off the constituents, given by $\Rq$ and $\Rqb$, and the interferences, encoded in $\mathcal{J}$. The singlet contribution in vacuum, which reads
\beq
\mathcal{R}^\text{vac}_\text{sing} = 2\omega^2\frac{p\cdot \bar p}{(p\cdot k)\, (\bar p\cdot k)} \,,
\eeq 
where $p$ ($\bar p$) is the four-momentum of the quark (antiquark), is the well-known antenna emission pattern \cite{Dokshitzer:1991wu}. It exhibits the characteristic feature of angular ordering. To set the notations for future use, let us also rewrite this spectrum as 
\beq
\label{eq:rsing-vac}
\mathcal{R}^\text{vac}_\text{sing} =  4\omega^2\frac{\delta \k^2}{\bkappa^2 \bbkappa^2} = 4\omega^2 \left( \frac{1}{\bkappa^2} + \frac{1}{\bbkappa^2} - 2\frac{\bkappa \cdot \bbkappa}{\bkappa^2 \bbkappa^2} \right) \,,
\eeq
where $\bkappa=\k-x\p$ (with $x = \omega/p^+$) is the gluon transverse momentum with respect to the quark (similarly, $\bbkappa = \k - \bar x \bar\p$ is with respect to the antiquark), and $\delta\k= \bkappa-\bbkappa$ is the relative transverse momentum of the $\qqb$-system. Note that the vector $\pip \equiv \delta \k /\omega$ is closely related to the opening angle of the pair, i.e., $|\pip| \simeq \sin \theta_\qqb$. The second relation in eq.~(\ref{eq:rsing-vac}) reflects directly the aforementioned decomposition of the spectrum into the independent components, given by the two former terms, and the interferences, described by the last term in eq.~(\ref{eq:rsing-vac}). For more details on the vacuum spectrum we refer the reader to the standard texts \cite{Dokshitzer:1991wu} and to our previous papers \cite{MehtarTani:2011tz, MehtarTani:2011gf}. Since the color octet spectrum is just a simple generalization related to conservation of color, see eq.~(\ref{eq:spec-general}), we will from now on only consider the singlet spectrum, postponing the discussion of an overall non-zero color charge to sec.~\ref{sec:octet}.

\begin{figure}
\centering
\includegraphics[width=0.3\textwidth]{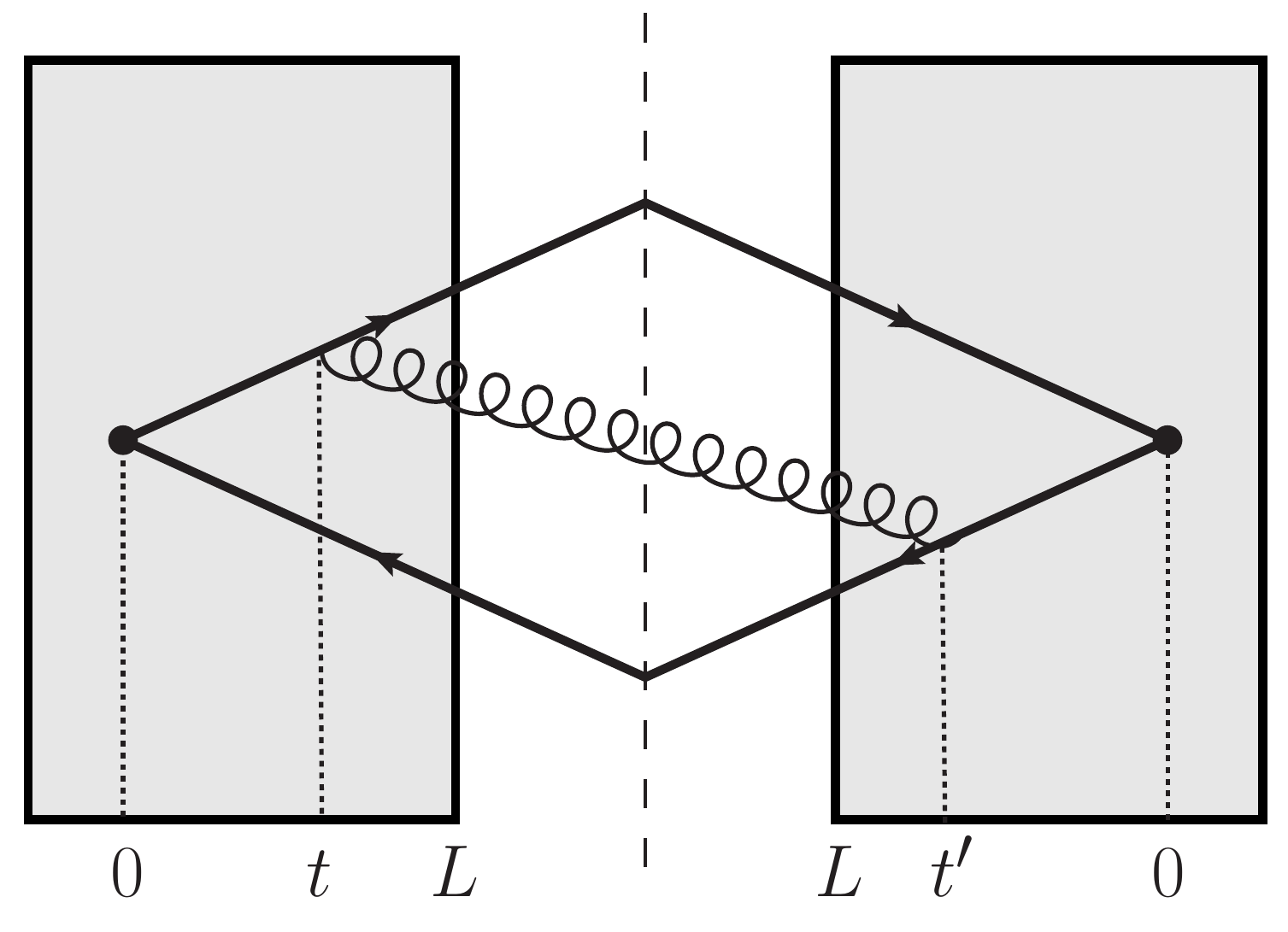} \quad
\includegraphics[width=0.3\textwidth]{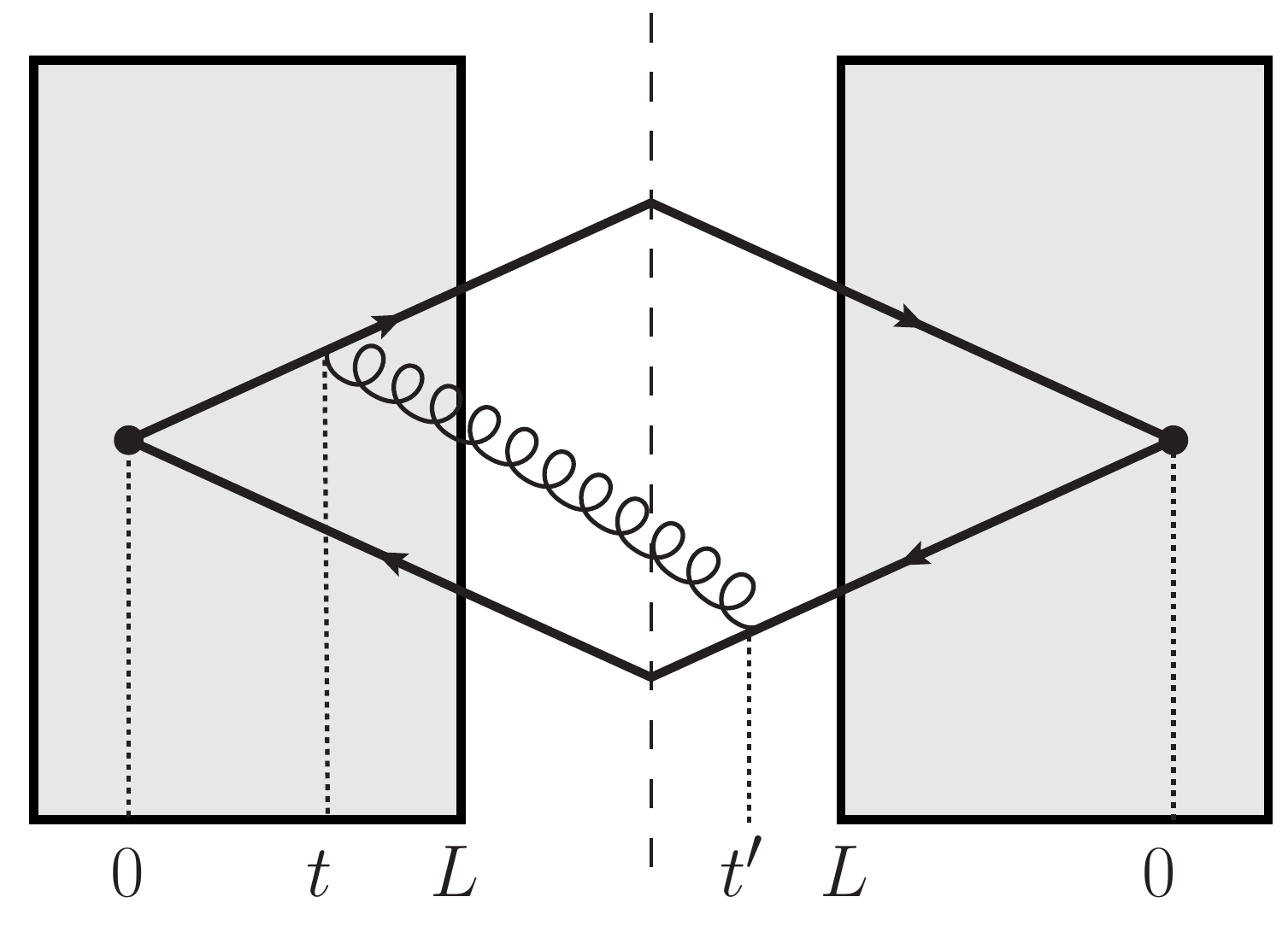}\quad
\includegraphics[width=0.3\textwidth]{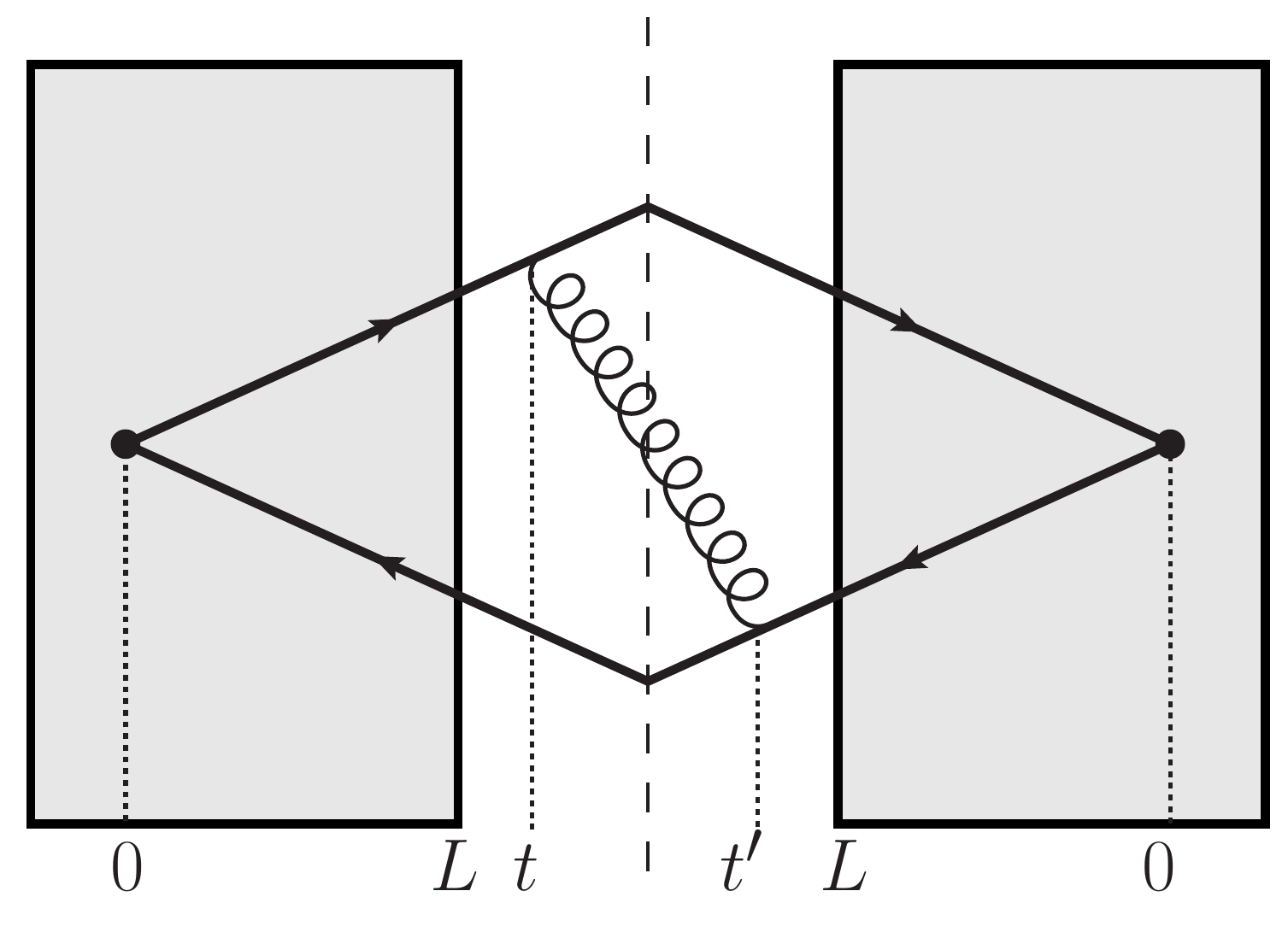}
\caption{The three parts of the gluon spectrum in the presence of a medium: ``in-in" component (left), ``in-out" component (center) and the ``out-out" vacuum component (right). The dashed line corresponds to the cut; on the left we depict the amplitude and on the right the complex conjugate amplitude.}
\label{fig:AntennaCrossSections}
\end{figure}

In the presence of a medium, both of the antenna constituents as well as the emitted gluon can interact with color charges in the quark-gluon plasma. The general setup is depicted in fig.~\ref{fig:AntennaKinematics}. We will mainly be interested in the most typical situation for studies of parton branchings in QCD, i.e., when the momenta of the projectiles is dominated by their positive light cone components. Interactions with the medium and the emission of the gluon are treated in the eikonal approximation which is valid at high energies up to the leading logarithm \cite{Baier:1996kr,Baier:1996sk}. Within this setup the interference spectrum was derived in \cite{MehtarTani:2011jw,CasalderreySolana:2011rz} and reads
\begin{align}
\label{eq:interferencespec-main}
\J&= \text{Re}  \int_0^\infty dt'\int_0^{t'} dt \big[1-\Delta_{\text{med}}(t) \big] \nn
&  \quad \times \int d^2\z \,\exp\left[-i \bkappa\cdot \z-\frac{1}{2}\int_{t'}^\infty d\xi\, n(\xi) \sigma(\z)+i\frac{\omega}{2}\delta\n^2 t\right] \nn
& \quad \times \left.  \left(\bdel_y-i\omega\,\delta\n\right)\cdot \bdel_z\,{\cal K}(t',\z\,;\,t, \y\,|\omega)\right|_{\y=\delta\n t} +\text{sym.}
\end{align}
where the symmetrical term is found by replacing $\bkappa \to \bbkappa$, which also implies $\pip \to -\pip$. The density of scattering centers as a function of time is given by $n(\xi)$. The independent components $\Rq$ and $\Rqb$ are found from eq.~(\ref{eq:interferencespec-main}) by taking $\pip \to 0$. For instance, the quark component is given by
\begin{align}
\label{eq:independentspec-main}
\Rq &= 2\text{Re}  \int_0^\infty dt'\int_0^{t'} dt \int d^2\z \,\exp\left[-i \k \cdot \z-\frac{1}{2}\int_{t'}^\infty d\xi\, n(\xi) \sigma(\z) \right] \nn
& \quad \times \left. \bdel_y \cdot \bdel_z\,{\cal K}(t',\z\,;\,t, \y\,|\omega)\right|_{\y=0} \,,
\end{align}
and an analogous expression holds for the antiquark component.\footnote{These expressions are precise modulo a formal cut-off prescription that is necessary in order to obtain the correct vacuum contribution \cite{Wiedemann:1999fq}.} The factor 2 in eq.~(\ref{eq:independentspec-main}) arises due to the possibility of rearranging the emission times in the amplitude and complex conjugate amplitude. In the simplest case of a medium with constant density the integration region of emission times in the amplitude and the complex conjugate amplitude is usually separated into three regions, namely
\beq
\label{eq:integration-regimes}
\int_0^\infty dt'\int_0^{t'} dt = \int_0^L dt'\int_0^{t'} dt +  \int_L^\infty dt'\int_0^{L} dt + \int_L^\infty dt'\int_L^{t'} dt \,,
\eeq
where $L$ denotes the length of the medium. The solutions for more involved medium profiles generalize this setup \cite{Baier:1998yf,Salgado:2003gb,Arnold:2008iy}. For obvious reasons, the three integration regions are called ``in-in", ``in-out" and ``out-out", respectively, see fig.~\ref{fig:AntennaCrossSections}.\footnote{This separation is somewhat artificial inasmuch as one only has two genuine types of radiation: a bremsstrahlung component due to the initial acceleration and a medium-induced component. We return to this point when discussing the ``in-out" component in sec.~\ref{sec:independent-inout}.} Let us currently describe the different components of the interference spectrum and, in parallel, the independent one in brief, postponing a more thorough analysis to secs.~\ref{sec:independent-analysis} and \ref{sec:interferences}.

The spectrum in eq.~(\ref{eq:interferencespec-main}) describes three distinct stages associated with the in-medium emission process. Initially one propagates the dipole consisting of the quark and antiquark, created in a color singlet state, from the the point where the $\gamma^\ast \to \qqb$ splitting occurred, chosen to be located at the origin of our coordinate system, to time $t$ when the emission of the gluon is initiated. At this stage, the interaction of the $\qqb$-pair with the medium is described by the survival probability \cite{MehtarTani:2011tz}
\beq
\label{eq:deltamed}
1-\Delta_\text{med}(t) = \exp\left[-\frac{1}{2} \int_0^{t} d\xi\, n(\xi)\ \sigma(\delta \n \, \xi)\right]  \,,
\eeq
where $\sigma(\pip \xi)$ is the dipole-medium forward scattering amplitude, see eq.~(\ref{eq:DipoleMedium-CrossSection}). Equation (\ref{eq:deltamed}) characterizes the rate of color decoherence of the pair prior to the emission due to color interactions with the medium. The function $\Delta_\text{med}(t)$ on the left hand side of eq.~(\ref{eq:deltamed}) is called the medium decoherence parameter \cite{MehtarTani:2010ma}. This factor determines a characteristic time-scale for decoherence of the $\qqb$-pair, denoted $\tdip$ and given by the solution of
\beq
\label{eq:decoh-time-general}
\frac{1}{2} \int_0^{\tdip} d\xi\, n(\xi)\ \sigma(\delta \n \, \xi) = 1 \,.
\eeq
Thus, for times $t \gg \tdip$ the color correlation of the pair is lost and the interference contribution is suppressed. Naturally, a corresponding process does not occur for a single particle in the medium since its own dipole moment is vanishing. Such a suppression factor is therefore not present in eq.~(\ref{eq:independentspec-main}). Correspondingly, for propagation  across the entire medium length one can define a characteristic opening angle $\thcrit$ by allowing the integration limit in eq.~(\ref{eq:decoh-time-general}) extend up to $L$ and solving for $|\delta \n|$.

The emission of the gluon, taking place at time $t$ in the amplitude and time $t'$ in the complex conjugate, is a common feature to both eq.~(\ref{eq:interferencespec-main}) and (\ref{eq:independentspec-main}). In this time interval, usually called the formation time, the dynamics of the current dipole, consisting of the gluon (in the amplitude) and quark (in the complex conjugate amplitude), is described by the path integral
\beq
\label{eq:k-path-integral}
\mathcal{K}\left(t',\z; t,\y| \omega \right) = \int\mathcal{D} [{\bs r}] \, \exp\left\{ \int_{t}^{t'} \!\!d\xi \left[ i\frac{\omega}{2} \dot{{\bs r}}^2(\xi) - \frac{1}{2} n(\xi) \sigma({\bs r}) \right] \right\} \,.
\eeq
It incorporates the Brownian motion of the gluon in the transverse plane from $t$ to $t'$, which comes from the accumulation of momentum from the medium, and results in a change of the dipole size from $\r(t)=\y$ to $\r(t')=\z$. The derivatives are simply conjugate to the transverse momenta at emission. However, the main distinction between the independent and interference spectra lies in the fact that in the former case the emissions and re-absorption (in the complex conjugate amplitude, that is) of the gluon takes place off the same antenna leg --- in the latter from two distinct ones. This gives rise to a different derivative structure for the two cases, cf. eqs.~(\ref{eq:interferencespec-main}) and (\ref{eq:independentspec-main}). Finally, one also has to include the separation of the transverse position of the emission points, which follows the quark and antiquark propagation in time as $\y = \pip \,t$, for the interference contribution.

After the formation of the gluon, the further propagation, from time $t'$ onward, is once more described by a survival probability, given by the second term in the second line of eq.~(\ref{eq:interferencespec-main}) and, correspondingly, by the last term in the first line of eq.~(\ref{eq:independentspec-main}). This final process takes into account the accumulation of additional momentum from the medium by the emitted gluon, see sec.~\ref{sec:independent-analysis} for further details.

The interference spectrum contains one further additional factor that is not present in for the independent radiation component, appearing in the last factor in the second line of eq.~(\ref{eq:interferencespec-main}). This phase factor can be written as $\delta \k^2 t/ (2\omega)$ is independent of the medium characteristics, see sec.~\ref{sec:interferences} for further details.

\section{A general picture: decoherence of QCD radiation}
\label{sec:discussion}

Before undertaking a technical discussion of the gluon spectrum in the presence of a medium contained in eq.~(\ref{eq:spec-general}), with the independent components and interferences given, respectively, by eqs.~(\ref{eq:independentspec-main}) and (\ref{eq:interferencespec-main}), it is worthwhile to consider the general characteristics of the chosen physical setup. At the outset we recall that the spectrum off a charge propagating through the medium, be it a single parton or an antenna, consists of two types of stimulated radiation:  a) the bremsstrahlung accompanying the creation (acceleration) of the charge, which we shall refer to as vacuum radiation, and b) the one that is stimulated directly by the medium interactions. Whereas  radiation of type a) has the typical broad, vacuum-like distribution, $\k^{-2}$, the induced radiation, of type b), is sustained only by some characteristic momentum transferred from the medium. We return to the physics of decoherence for a single charge in sec.~\ref{sec:independent-analysis}.

In the absence of a medium only the former, vacuum radiation component is at work. For the antenna, gluons are only radiated at relatively small emission angles $\theta < \theta_\qqb$, where they are produced independently by each of the antenna constituents. This comes about due to the fact that the gluon transverse wave length must  in this case be smaller than the dipole size when the gluon is formed in order to resolve the internal structure of the pair. In the opposite case, $\theta > \theta_\qqb$, the spectrum is suppressed since large angle gluon emission is sensitive to the total color charge of the pair (which in the singlet case is zero). This suppression is achieved through destructive interferences so that the color flow is restricted to a cone with aperture given by the opening angle of the pair.

We can further deepen our understanding of this phenomenon by considering the relevant scales involved in the problem. In vacuum, the characteristic hard scale, which necessarily is related to the opening angle of the pair, is simply $\Qhard = \delta \k$. It determines the maximal transverse momentum of the gluons that can be produced by the system. In fact, for gluon transverse momenta $\k \gg \delta \k$ the spectrum is power-suppressed, $\sim \k^{-4}$, see eq.~(\ref{eq:rsing-vac}). Conversely, considering the emission off one lone charge in the vacuum does not involve any hard scale --- therefore no cut-off modifies the transverse spectrum, which remains vacuum-like $\sim \k^{-2}$ in the whole phase space.

In the general case we have to consider the additional transverse dynamics induced by the interactions with the medium. We follow closely the discussion in \cite{MehtarTani:2011gf}, where it was realized that the gross features of the antenna spectrum in a medium can be understood in terms of two characteristic scales.\footnote{We would like to thank A.~H. Mueller for inspiring this interpretation.} These are easily obtained from an analysis of the medium decoherence parameter defined in eq.~(\ref{eq:deltamed}), which we presently conjecture to be in the following form
\beq
\label{eq:deltamed-approx}
1-\Delta_\text{med} \sim e^{- \,\Qmed^2 \,r_\perp^2 } \,,
\eeq
postponing the derivation and detailed discussion to sec.~\ref{sec:ho-approximation}, also see \cite{MehtarTani:2011tz}. First and foremost one has the transverse color screening length of the medium. It is given by the inverse of the saturation scale of the medium, denoted by $\Qmed$. In a dense medium it is given by $\Qmed^2 = \sqrt{\hat q L}$, where $\hat q$ is the medium transport coefficient and $L$ is the size of the medium. The other relevant length scale of the problem is simply related to the maximal antenna size in the medium, namely $r_\perp=\theta_{q\bar q} L$. We discuss these features in more technical detail in sec.~\ref{sec:interferences}.

The resulting spectrum is characterized by the competition between the two medium-induced scales and the intrinsic antenna scale. The largest transverse scale determines the hard scale of the problem, or
\beq
Q_\text{hard} = \max (\Qdip, \Qmed, |\delta \k|)\,,
\eeq
which assigns a maximal transverse momentum that can be generated by the system under consideration. Below, for the sake of argument, we will always assume that the vacuum scale $|\delta \k|$ is much smaller than the medium scales and we will only consider radiation at angles larger than the pair opening angle, unless stated otherwise. This is the relevant regime for studying interferences, as radiation taking place inside the pair always is independent.

The former regime, defined by $Q_\text{hard} = \Qdip$, will be denoted the ``dipole" regime since the intrinsic dipole scale governs the dynamics of the radiative processes. The second regime will, on the other hand, be denoted the ``decoherence" regime.  In this case the dynamics are dominated by scales induced by the medium since $Q_\text{hard} = \Qmed$. 

In terms of angles, the characteristic momentum scale translates accordingly into a maximal angle of gluon emission given by $\thmed = Q_\text{hard}/\omega$. Note that this angle becomes arbitrary large in the infrared limit \cite{MehtarTani:2011tz}. For $\thmed > \theta_\qqb$ this implies that the color flow has been reshuffled to higher angles and color conservation is only achieved at $\theta > \thmed$. In this regime, the overall physical picture is then analogous to the vacuum case in the sense that the pair opening angle is replaced by $\max(\thmed,\theta_\qqb)$ as the angle that governs the transition between independent emissions at `small' angles and emissions off the total color charge at `large' angles (which in the color singlet case is strongly suppressed, as expected). 

The considerations above define two distinct regimes for in-medium gluon emissions, which we proceed to discuss in some detail. A summary of relevant scales have been collected in tab.~\ref{tab:regimes}.
\begin{table}[t]
\begin{center}
\begin{tabular}{l|c|c}
 & ``dipole" regime & ``decoherence" regime\\
\hline
hard scale ($Q_\text{hard}$) & $\Qdip$ & $\Qmed$ \\
critical opening angle ($\thcrit$) & $\theta_\qqb \ll \thcrit$ & $\theta_\qqb \gg \thcrit$ \\
decoherence time ($\tdip$) & $\tdip \gg L$ & $\tdip \ll L$\\
max. emission angle ($\theta_\text{max}$) & $(\omega r_\perp)^{-1}$ & $\Qmed/\omega$ \\
\end{tabular}
\caption{\label{tab:regimes} Overview of the characteristics governing emission off the antenna in medium in the two possible situations when $\Qdip, \Qmed > |\delta \k|$.}
\end{center}
\end{table}%

\subsection{The ``dipole" regime (``partial decoherence"): $r_\perp < \Qmed^{-1}$}
\label{sec:discussion-dipole}

The ``dipole" regime applies to situations when the size of the antenna during the passage through the plasma always is smaller than the transverse screening length, so that the condition $r_\perp < \Qmed^{-1}$ is fulfilled. This happens because the characteristic decoherence time, defined in eq.~(\ref{eq:decoh-time}), is much larger than the interaction length, $\tdip \gg L$. Analogously, one can also visualize that the opening angle of the pair is smaller than some critical angle, which we denote $\thcrit$ in tab.~\ref{tab:regimes}. This angle can be deduced from the general definition of the decoherence parameter, see comment below eq.~(\ref{eq:decoh-time}), and in the approximation used throughout, cf. sec.~\ref{sec:ho-approximation}, this angle becomes simply $\thcrit = 1/\sqrt{\hat q L^3}$. The $\qqb$-pair thus preserves its color correlation throughout the passage and, in this sense, probes the medium structure on short transverse distances.

Accordingly, the hardest scale of the problem is $\Qdip$, determining the characteristic cut-off of the transverse momentum spectrum. Note that in this case the medium-induced gluon spectrum is directly proportional to the decoherence parameter which reads
\beq
\Delta_\text{med} = \frac{1}{12} \Qmed^2 r_\perp^2 \,,
\eeq
after expanding the right hand side of eq.~(\ref{eq:deltamed-approx}). The cross-section reveals the characteristic color-transparency behavior and results in a partial decoherence of the antenna spectrum \cite{MehtarTani:2010ma,MehtarTani:2011gf}. The emitted gluons have relatively long formation times and therefore are not strongly affected by the broadening in the medium. We postpone the discussion of further technical details relevant for this regime to sec.~\ref{sec:interferences-dipole}.

\subsection{The ``decoherence" regime: $r_\perp > \Qmed^{-1}$}
\label{sec:discussion-decoherence}

In this regime, the hardest scale of the problem is given by $\Qmed$ and the quark and the antiquark are correlated only for a very short time after entering the medium, i.e., for times $\tdip < L$. After complete decoherence is achieved the constituents are free to radiate independently along the remaining path-length through the medium. In this sense, the medium probes the internal structure of the antenna. Therefore, as the decoherence parameter is saturated at the maximal value, i.e., 
\beq
\Delta_\text{med} = 1 \,,
\eeq
the interferences are suppressed and all gluons with $k_\perp < \Qmed$ are emitted incoherently. As in the previous case, transverse momenta $k_\perp > \Qmed$ are not allowed by the medium-induced radiation hence restoring color coherence at angles $\theta > \thmed = \Qmed/\omega$. So long as $\thmed > \theta_\qqb$ the spectrum has lost all information about the opening angle of the pair --- this defines total decoherence. Indeed, the angle $\thmed$ acts as a new `opening angle' of the pair.

We discuss the detailed characteristics of the independent component and the interferences for the ``decoherence" regime in secs.~\ref{sec:independent-analysis} and \ref{sec:interferences-dense}, respectively.

\section{The antenna spectrum in the harmonic oscillator approximation}
\label{sec:ho-approximation}

The details of the spectrum in eqs.~(\ref{eq:interferencespec-main}) and (\ref{eq:independentspec-main}) are determined by the dipole-medium cross section $\sigma$ which contains the dynamics of the interaction. As described above, precisely what dipole is probed by the medium at a given time changes in course of the process. The general expression for $\sigma(\r)$ reads
\beq
\label{eq:DipoleMedium-CrossSection}
\sigma (\r) = \int \frac{d^2\q}{(2\pi)^2} \mathcal{V}^2(\q) \big[1 - \cos( \r \cdot \q)  \big] \,,
\eeq
where $\mathcal{V}(\q)$ is the medium potential or, in other words, the elastic scattering rate. The details of the nature of the medium are in our context not of great importance, noticing that at leading-logarithmic approximation the cross section simply is proportional to $\r^2$. At this level, we can capture the universal features of the multiple scatterings and examine the spectrum analytically. Summing over multiple scattering with the medium lead to the exponentiation of the product of $\sigma (\r)$ and the medium density, $n(t)$. The harmonic oscillator approximation consists of assuming that this product simply reads
\beq
n(t) \sigma(\r) \approx \frac{1}{2} \hat q(t) \r^2 \,.
\eeq
This approximation is strictly valid for multiple soft scattering in a 'dense' medium characterized by the condition that the gluon formation time is much larger than the mean free path, i.e., $\tform \gg \lambda_\text{mfp}$, and breaks down when the projectile becomes sensitive to isolated scattering centers, i.e., for small dipoles such that $\r^{-1} \gg \int dt \,\hat q (t)$. This suggests in turn that the approximation is valid for soft gluon production and breaks down in the hard sector where one is more sensitive to hard, atypical medium interactions, see, e.g., \cite{CaronHuot:2010bp,Zapp:2012nw}. In the latter case, an analysis incorporating the microscopical structure of the medium, such as done, e.g., in \cite{MehtarTani:2011gf}, is more relevant. In the harmonic oscillator approximation there is also a direct relation between the medium transport parameter $\hat q$ and the transverse momentum accumulated by a particle traversing the medium \cite{Baier:1996sk}, which suggests that it can be interpreted as the momentum broadening per unit length. In line with previous considerations, we assume for the time being a constant medium density from the initial production point to some length $L$, such that $\hat q(t) = \hat q \Theta(L-t)$.

In the harmonic approximation the survival factors become Gaussian functions. In particular, the decoherence parameter simply reads
\beq
\label{eq:decoh-parameter}
\Delta_{\text{med}}(t) =1-\exp\left[-\frac{1}{12}\hat q\, \delta\n^2\, t^3\right] \,,
\eeq
where the dipole considered is the one related to the $\qqb$ opening angle, i.e., $\r^2(t) =( \pip \,t)^2$. Here we point to the fact that the size grows linearly with time. We identify the decoherence time scale, defined in eq.~(\ref{eq:decoh-time-general}), as
\beq
\label{eq:decoh-time}
\tdip = \big(\hat q\theta^2_\qqb \big)^{-1/3} \,.
\eeq
The same logic governs the process of gluon formation, described by eq.~(\ref{eq:k-path-integral}). For this situation, however, the size of the dipole fluctuates around a characteristic size. As a first heuristic estimate, let us invoke a typical Brownian diffusion process: during the typical formation time, which involves a certain number $N_\text{coh}$ of coherent scatterings, $\tform = \lambda_\text{mfp} N_\text{coh}$, the gluon accumulates a transverse momentum $\kform^2 = m_D^2 N_\text{coh}$, where $m_D$ is the so-called Debye mass which constitutes the typical momentum transfer from the medium in a single scattering. This gives immediately an estimate of the typical formation time of an induced gluon, which reads
\beq
\label{eq:formation-time}
\tform = \sqrt{\frac{\omega}{\hat q}}\,,
\eeq
where the medium transport parameter becomes $\hat q = \sqrt{m_D^2/\lambda}$ for the purpose of this heuristic estimate. This differs from the formation time in vacuum, which goes like $\tform = 1/(\omega \theta^2)$ and is typically long for soft gluons at a fixed angle $\theta$. In contrast, soft gluons are induced quite rapidly in the medium. It follows that the accumulated momentum becomes
\beq
\label{eq:formation-scale}
\kform^2 = \sqrt{\omega \hat q} \,,
\eeq
and the typical dipole size governing eq.~(\ref{eq:k-path-integral}) is therefore $\r^2 = \kform^{-2}$. Since this is the size of the quark-gluon dipole at formation time, $\r^2(\tform) = (\thform \tform)^2$, we deduce the characteristic angle for these emissions to be $\thform = (\hat q / \omega^3)^{1/4}$. Thus, in the same sense as decoherence of the $\qqb$-pair emerges at $t > \tdip$ the gluon becomes decorrelated from the quark at times $t > \tform$. 

For a more precise treatment we note that in this scheme the Green's function in eq.~(\ref{eq:k-path-integral}) is that of a harmonic oscillator and allows an exact, analytical solution. This is given by
\beq
{\cal K}(t',\z\,;\,t, \y\,| \omega)=\frac{A}{\pi i}\, \exp\left[iAB(\z^2+\y^2)-2iA\,\z\cdot\y\right],
\eeq
where, in a constant medium presently under consideration,
\beq
\label{eq:ab-coefficients}
A=\frac{\omega \Omega}{2\sin(\Omega\Delta t)}\,, \qquad B=\cos(\Omega\Delta t ) \,,
\eeq
with $\Delta t = t' - t$ and the parameter $\Omega^{-1} = (1+i)\, \tform$ is roughly the inverse of the formation time, as estimated in eq.~(\ref{eq:formation-time}). The functions $A, B$ in the case of a smooth medium profile or for an expanding medium are, e.g., given in \cite{Baier:1998yf,Salgado:2003gb,Arnold:2008iy}. Within the harmonic oscillator approximation the spectrum in eq.~(\ref{eq:interferencespec-main}) can be solved in coordinate-space representation. This simplifies the numerical analysis which we will return to in sec.~\ref{sec:numerics} and we therefore detail the corresponding expressions in appendix~\ref{sec:HOcoordinate}. Presently, we will pursue another strategy which makes a more analytical treatment of the spectrum possible.

This can be achieved in a mixed representation involving the longitudinal position of gluon emission and its corresponding transverse momentum. Taking the Fourier transform of the transverse endpoint of the path integral,
\beq
\K(t',\z; t, \y | \omega) = \int \frac{d^2 k'}{(2\pi)^2} e^{i \k' \cdot \z}  \tilde \K(t', \k' ; t,\y |\omega) \,, 
\eeq
where 
\beq
\tilde \K(t', \k' ; t,\y |\omega) = \frac{1}{B} \exp \left[-i \frac{\k'^2}{4AB} + i \frac{A \left( B^2 -1 \right)}{B} \y^2 - i \frac{\k' \cdot \y}{B} \right] \,,
\eeq
is the propagator in a mixed representation. Then we find that the ``\inin" component of eq.~(\ref{eq:interferencespec-main}) becomes
\begin{align}
\label{eq:j-inin-1}
\J^\text{in-in} & = \text{Re} \int_0^L dt' \int_0^{t'} dt \int\frac{d^2\k'}{(2\pi)^2} \, \big[1 - \Delta_\text{med}(t) \big] \mathcal{P}(\bkappa - \k', L-t') \nn
&\quad \times \frac{1}{B^2} \k' \cdot \left\{ \k' - B \left[1 + \frac{2A(B^2-1) t}{\omega B} \right] \omega \pip \right\} \nn
&\quad \times \exp \left[ i\frac{\omega \,\pip^2 \,t}{2}  - i \frac{\k'^2}{4AB} - i \frac{\k' \cdot \pip \, t}{B} + i \frac{A(B^2-1) \, \pip^2 t^2}{B} \right] \,+\, \text{sym.} \,,
\end{align}
where the coefficients $A$ and $B$ are given in eq.~(\ref{eq:ab-coefficients}). In eq.~(\ref{eq:j-inin-1}), we have defined the function
\beq
\label{eq:p-broadening}
\mathcal{P}(\k, \xi) \equiv  \frac{4\pi}{\hat q \xi} \exp \left[- \frac{\k^2}{\hat q \xi} \right] \,,
\eeq
which describes the probability of having accumulated a certain transverse momentum squared $\k^2$ after traversing the longitudinal distance $\xi$. This factor emerges due to the classical transverse momentum broadening of the gluon in the wake of the quantum emission process. The presence of this factor corresponds to the reshuffling of gluon momenta, and does not contribute to the absolute yield of produced gluons, since $\int \frac{d^2\k}{(2\pi)^2}\mathcal{P} (\k,\xi) =1$. We demonstrate the role of this particular effect numerically in some more detail in appendix~\ref{sec:broadening}. 

Naturally, this final-state broadening is absent for the remaining ``\inout" and ``\outout" components of eq.~(\ref{eq:interferencespec-main}). Within the harmonic oscillator approximation, the vacuum-medium interference term reads
\begin{align}
\label{eq:j-inout-1}
\J^\text{in-out} & = - \text{Re} \int_0^L dt \, \big[1 - \Delta_\text{med}(t) \big] \frac{2i \, \omega}{\tilde B^2 \, \bkappa^2}\,  \bkappa \cdot \left\{ \bkappa - \tilde B \left[1 + \frac{2 \tilde A( \tilde B^2-1) t}{\omega \tilde B} \right] \omega \pip \right\} \nn
&\quad \times \exp \left[ i\frac{\omega \,\pip^2 \,t}{2}  - i \frac{\bkappa^2}{4 \tilde A \tilde B} - i \frac{\bkappa \cdot \pip \, t}{\tilde B} + i \frac{\tilde A(\tilde B^2-1) \, \pip^2 t^2}{\tilde B} \right]  \,+\, \text{sym.} \,,
\end{align}
where, in this case,
\beq
\label{eq:tildeab-coefficients}
\tilde A = \frac{\omega \Omega}{2\sin[\Omega(L- t)]}\,, \qquad \tilde B = \cos[ \Omega (L-t )] \,.
\eeq
Finally, emissions taking place outside of the medium are given by the ``out-out" component, which reads
\beq
\label{eq:j-outout-1}
\mathcal{J}^\text{out-out} =  \big[ 1 - \Delta_\text{med}(L) \big] \frac{4\omega^2 \, \bkappa \cdot \bbkappa}{\bkappa^2 \bbkappa^2} \cos \left[ \frac{(\bkappa + \bbkappa) \cdot \r}{2} \right] \,,
\eeq
where $\r = \pip L$. In particular, the term in eq.~(\ref{eq:j-outout-1}) is responsible for the decoherence of the vacuum radiation. Note that it is proportional to the vacuum interference emission pattern, cf. the last term in eq.~(\ref{eq:rsing-vac}), which only contributes outside of the opening angle. This term involves only the interaction of the quark and antiquark with the medium and is the leading term of the spectrum in the infrared limit \cite{MehtarTani:2011tz}. Emissions beyond the infrared limit must in addition take into account gluon rescattering and involve also the ``\inin" and ``\inout" contributions --- note that this also holds in the vacuum, see the section below. Before examining the complicated features of the interferences in medium we will first examine the independent spectrum off a single charge.

\subsection{Interferences in vacuum}
\label{sec:interferences-vacuum}

As a brief interlude, we would like to point out some peculiar features of the interference spectrum that arise in vacuum. Above it was noted that in this case the interferences are not simply given by the term describing emissions outside of the medium but involve all three components $\mathcal{J}^\text{in-in}$, $\mathcal{J}^\text{in-out}$ and $\mathcal{J}^\text{out-out}$. This can be seen, for instance, taking the $\hat q \to 0$ limit. In this case,
\begin{align}
\label{eq:j-inin-vacuum}
\mathcal{J}^\text{in-in} &= 4\omega^2 \frac{\bkappa \cdot \bar \bkappa}{\bkappa^2 \bar \bkappa^2} \left\{1 - \cos \left(\frac{\bkappa^2}{2\omega}L \right)  - \cos \left(\frac{\bar \bkappa^2}{2\omega}L \right) + \cos \left[\frac{(\bkappa + \bar \bkappa)\cdot \r}{2} \right] \right\} \,, \\
\label{eq:j-inout-vacuum}
\mathcal{J}^\text{in-out} &= 4\omega^2 \frac{\bkappa \cdot \bar \bkappa}{\bkappa^2 \bar \bkappa^2} \left\{\cos \left(\frac{\bkappa^2}{2\omega}L \right)  + \cos \left(\frac{\bar \bkappa^2}{2\omega}L \right) -2 \cos \left[\frac{(\bkappa + \bar \bkappa)\cdot \r}{2} \right] \right\} \,, \\
\label{eq:j-outout-vacuum}
\mathcal{J}^\text{out-out} &=  4\omega^2 \frac{\bkappa \cdot \bbkappa}{\bkappa^2 \bbkappa^2} \cos \left[ \frac{(\bkappa + \bbkappa) \cdot \r}{2} \right] \,,
\end{align}
such that
\beq
\label{eq:j-total-vacuum}
\mathcal{J} \equiv \mathcal{J}^\text{in-in} + \mathcal{J}^\text{in-out} + \mathcal{J}^\text{out-out} = 4\omega^2 \frac{\bkappa \cdot \bbkappa}{\bkappa^2 \bbkappa^2} \,,
\eeq
in accordance with the last term in eq.~\eqref{eq:rsing-vac}. In other words, the presence of a boundary, in our case delimited by $L$, induces a `delay effect' causing the distortion of the interference effects. The genuine vacuum contribution, i.e., the one that does not depend on $L$, can be found in $\mathcal{J}^\text{in-in}$, see the first term in eq.~(\ref{eq:j-inin-vacuum}). Taking additionally the $L \to 0$ limit in eqs. (\ref{eq:j-inin-vacuum})-(\ref{eq:j-outout-vacuum}), we are only left with $\mathcal{J}^\text{out-out}$ contribution, as expected. As already pointed out above, this `delay effect' is irrelevant for soft gluon production.

\section{The single-inclusive independent spectrum revisited}
\label{sec:independent-analysis}

From the heuristic analysis of the behavior of the survival probabilities we have found that the medium induces gluons with a preferred transverse momentum, $\k^2(\tform) \sim \kform^2$, see eq.~(\ref{eq:formation-scale}). Considering long formation times, $\tform \sim L$, we find the maximal transverse momentum that can be accumulated to be given by
\beq
\Qmed = \sqrt{\hat q L} \,,
\eeq
which is the scale that naturally governs the momentum broadening, see eq.~(\ref{eq:p-broadening}). However, we will shortly demonstrate that the spectrum is mainly dominated by the emissions of gluons with short formation times. Reverting for a moment to heuristic arguments, this fact becomes clear when realizing that the spectrum of gluons will be proportional to the number of available production lengths, given in the simplest case by $\tform \sim \lambda_\text{mfp} N_\text{coh}$, for gluon production, i.e.
\beq
\label{eq:bdmps-estimate}
\omega \frac{dN}{d\omega} \sim \alpha_s \frac{L}{\tform} \sim \alpha_s \sqrt{ \frac{\omega_c}{\omega} } \,,
\eeq
where $\omega_c = \hat q L^2$ is a characteristic gluon energy. Thus, gluons with short formation times, $\tform \ll L$, can be produced anywhere along the medium length. In the opposite case, $\tform \gg L$ (implying in turn $\omega \gg \omega_c$), the estimate of eq.~(\ref{eq:bdmps-estimate}) breaks down and the spectrum is strongly suppressed. This is the so-called Landau-Pomeranchuk-Migdal (LPM) suppression in QCD.

Additionally, we have to keep track of the bremsstrahlung. Usually, one simply subtracts a pure vacuum component,
\beq
\left. \mathcal{J} \right|_{\delta \n \to 0} = \frac{4 \omega^2}{\k^2} \,,
\eeq
see eq.~(\ref{eq:j-total-vacuum}), from the total spectrum to obtain a purely medium-induced quantity. But, as already illustrated for a fixed, non-zero angle $|\delta \n|$ in sec.~\ref{sec:independent-analysis}, this contribution is in reality built up from both early and late time emissions. We will therefore keep track of this component since it is imperative in order to better understand the role of the contribution of bremsstrahlung in the medium. 

In the following subsections we will go into the analytical details of how both of these types of emissions are generated. Finally, in subsection \ref{sec:indep-revisited}, we show how the main features of the independent spectrum can be obtained performing some simple approximations that capture the dynamics of the process. The insights gained along the way will help us to identify the main aspects governing the interference contribution.

\subsection{The ``in-in" component}
\label{sec:independent-inin}

The independent ``in-in" component can be found by taking the $\pip \to 0$ limit in eq.~(\ref{eq:j-inin-1}), and reads
\begin{multline}
\label{eq:r-in-in-1}
\Rq^\text{in-in} = 2 \text{Re} \int_0^L dt' \int_0^{t'} dt \int\frac{d^2\k'}{(2\pi)^2} \mathcal{P}(\k-\k',L-t') \frac{\k'^2}{\cos^2\left(\Omega \Delta t \right)} \exp \left[- i \frac{\k'^2}{2 \omega \Omega}\tan \left( \Omega \Delta t\right) \right] \,.
\end{multline}
Noting that the integrand of eq.~(\ref{eq:r-in-in-1}) is a total derivative, can perform the integral over $t$ to obtain
\beq
\label{eq:r-in-in-2}
\Rq^\text{in-in} = 2  \text{Re} \int_0^L dt'  \int\frac{d^2\k'}{(2\pi)^2}\P(\k-\k',L-t') \,2i\omega \exp \left[ (1-i) \frac{\k'^2}{2 \kform^2} \tan(\Omega t')\right] \,,
\eeq
where we have dropped a purely imaginary term in the last step. 

The argument of the tangent in eq.~(\ref{eq:r-in-in-2}), which goes roughly like $t'/\tform$, defines two regimes, namely i) late emissions, $t' \gg \tform$, and ii) early emissions, $t' \ll \tform$, giving rise to two distinct physical processes. To separate these we introduce a spurious parameter, a real number $c \gg 1$ such that $c \tform \ll L$, which eventually drops out from any physical quantity. We proceed to analyze the spectrum in the two cases.

\subsubsection{Late emissions: $t' \gg \tform$}
\label{sec:independent-inin-late}

For emissions that happen a long time after the creation point, $t' > c\tform$, we can approximate $\tan(\Omega\, t') \approx -i$. Hence, eq. (\ref{eq:r-in-in-2}) yields
\beq
\label{eq:r-in-in-3}
\left. \Rq^\text{in-in}\right|_{t'>c\tform} =4\omega  \int_{c\tform}^L dt'  \int\frac{d^2\k'}{(2\pi)^2}\P(\k-\k',L-t')\sin\left(\frac{\k'^2}{2 \kform^2}\right)e^{-\frac{\k'^2}{2 \kform^2}}\,.
\eeq
As indicated above, the single-gluon spectrum can be interpreted as a two-step process. The latter two factors in eq.~(\ref{eq:r-in-in-3}) describe the quantum emission of a gluon with momentum $\k'$, which occurs approximately at time $t'$, while the function $\mathcal{P}(\k-\k',L-t')$ describes the subsequent classical, Brownian motion of the gluon along the remaining path through the medium. The emission spectrum peaks around the characteristic momentum $\kform^2$, and is exponentially suppressed for $\k'^2 > k^2_f$. However, the final-state momentum broadening smears the distribution according to the saturation scale of the medium, given roughly by $\Qmed$.

\subsubsection{ Early emissions: $ t'< \tform$}
\label{sec:independent-inin-early}

If the emission happens early during the passage through the medium, $t' < c\tform$, the $t'$ dependence in the broadening probability $\P$ can be neglected. Then the remaining terms in eq.~(\ref{eq:r-in-in-2}) simplify to
\beq
\label{eq:r-in-in-4}
\left. \Rq^\text{in-in} \right|_{t'<c\tform} =  \text{Re} \int\frac{d^2\k'}{(2\pi)^2}\P(\k-\k',L) \frac{4i \omega}{\Omega}\int_0^{\beta} dx \exp \left[ (1-i) \frac{\k'^2}{2 \kform^2} \tan(\Omega t') \right] \,,
\eeq
where $\beta \equiv (1-i) c/2$. The integral over the time can now be performed exactly, since
\beq
\label{eq:exp-integral-definition}
\int dx \,e^{a \tan x} = \frac{i}{2}\left\{ e^{ia}E_1 \left[a (i-\tan x) \right] - e^{-ia} E_1\left[-a (i + \tan x) \right] \right\} \,+\text{const.}
\eeq
where $E_1(x)= - Ei(-x)$ is the exponential integral. Keeping only the leading term in $c$, such that $\tan(\beta)\approx -i + 2ie^{-2i \beta}$, we find that
\begin{align}
\label{eq:exp-integral}
\int_0^{\beta} dx \exp \left[ (1-i) \frac{\k'^2}{2 \kform^2} \tan(\Omega t')\right]  \simeq &  -\frac{i}{2} \Big\{ e^{ia}\big[ E_1(ia)-E_1(2ia) \big] \nn
& \quad -e^{-ia} \big[ E_1(-ia) + \gamma + \ln(-2ia) - 2i \beta \big] \Big\},
\end{align}
where now $a \equiv (1-i) \k'^2/(2 \kform^2)$ and we have expanded the last term using $E_1(x)\simeq -\gamma-\ln(x)$ for $x \ll1$ (here, $\gamma = 0.577...$ is the Euler-Mascheroni constant). The last term in eq.~(\ref{eq:exp-integral}) gives rise to a term of the form of the eq.~(\ref{eq:r-in-in-3}), however with $t'$ running from 0 to $c\tform$. Adding this to eq.~(\ref{eq:r-in-in-3}) removes the spurious dependence on $c$, as anticipated.

The remaining pieces of eq.~(\ref{eq:exp-integral}) are also not proportional to $L$ and can be dropped. We note that taking $\k'^2 > \kform^2$ in eq.~(\ref{eq:exp-integral}) we can gather the leading terms of the exponential integral at large arguments, i.e., $E_1(x) \approx e^{-x}/x$, to obtain
\beq
\label{eq:r-in-in-short-leading}
 \frac{4i \omega}{\Omega}\left. \int_0^{\beta} dx \exp \left[ (1-i) \frac{\k'^2}{2 \kform^2} \tan(\Omega t')\right]  \right|_{\k'^2 > \kform^2} \simeq \frac{8\omega^2}{\k'^2} \,,
\eeq
in addition to sub-leading terms that are suppressed by $\exp[- \k'^2/(2\kform^2)]$ and will be neglected in the following. Thus, the actual contribution from early emission seem to generate a hard vacuum-like component at $\k'^2 > \kform^2$, which arises from emission at a very early stage in the medium, followed by the rescattering of the emitted gluon in the medium. Note, however, that the collinear divergence expected for such a contribution is naturally cut off by construction in  eq.~(\ref{eq:exp-integral}). The momentum broadening aside, this contribution is twice larger than expected for a genuine bremsstrahlung emission. Now recall that in the presence of broadening the hardest scale of the problem automatically becomes $\Qmed$.  Thus, in the regime of hard, final-state transverse momenta $\k^2 > \Qmed^2$, where the effects of broadening can be neglected, $\mathcal{P}(\k-\k',L)|_{\k^2 > Q_s^2} \simeq (2\pi)^2\delta(\k-\k')$, the resulting ``in-in" contribution ($\sim8 \omega^2/\k^2$) is balanced out by the ``in-out" and ``out-out" contributions to restore a genuine vacuum contribution $\sim 4 \omega^2/\k^2$, also see the following subsection. The mismatch between the in-medium and out-of-medium bremsstrahlung contributions for momenta $\kform < k_\perp < \Qmed$ is an interesting feature which deserves further study. 

In sec.~\ref{sec:interferences-dense} we will show how interferences are crucial in this particular regime establishing color coherence.

\subsection{The ``in-out" and ``out-out" components}
\label{sec:independent-inout}

According to our decomposition in eq.~\eqref{eq:integration-regimes} the ``in-out" and ``out-out" components describe mainly radiation taking place either close to the boundary of the medium or outside of it. Formally, for the general case of smooth medium profiles we cannot formally distinguish between them, see, e.g., \cite{Arnold:2008iy}.

Taking the limit of vanishing opening angle, $\pip \to 0$, in eq.~(\ref{eq:j-inout-1}) we find the independent ``in-out" component,
\beq
\label{eq:r-in-out-1}
\Rq^\text{in-out} = - 2\text{Re} \int_0^L dt \, \frac{2i \, \omega}{\cos^2\left[\Omega (L-t) \right]} \exp \left\{- i \frac{\k^2}{2\omega \Omega} \tan \left[\Omega (L-t) \right] \right\} \,,
\eeq
which accounts for interference effects between emissions inside and outside the medium. From the discussion in the previous section, it becomes clear that this contribution is only supported in the final formation-time slice away from the medium border, see below. After integrating over the emission time $t$ we find
\beq
\Rq^\text{in-out} = - \frac{8\omega^2}{\k^2}\text{Re} \left\{1-\exp\left[ (1-i)\frac{\k^2}{2 \kform^2}\tan\Omega L \right] \right\}\,.
\eeq
Recalling that $|\Omega|L\gg 1$ as long as we consider $\tform \ll L$, we obtain
\beq
\label{eq:r-in-out-2}
\left. \Rq^\text{in-out}\right|_{\tform \ll L} = - \frac{8\omega^2}{\k^2}\left[1-\cos\left(\frac{\k^2}{2 \kform^2}\right)e^{-\frac{\k^2}{2 \kform^2}}\right] \,.
\eeq
Finally, the ``out-out" component is given by
\beq
\label{eq:r-out-out-1}
\Rq^\text{out-out} = \frac{4\omega^2}{\k^2} \,,
\eeq
where we have utilized a regularization of the time integral at infinity as described in \cite{Wiedemann:1999fq}. Summarizing the leading behavior of the out-of-medium contributions,
\beq
\label{eq:r-out-of-medium}
\mathcal{R}_q^\text{in-out} + \mathcal{R}_q^\text{out-out} = \begin{dcases}  \frac{4\omega^2}{\k^2} & \text{for  }\k^2 < \kform^2\,, \\ - \frac{4\omega^2}{\k^2} & \text{for  } \k^2 > \kform^2 \,,\end{dcases}
\eeq
we notice the change of sign that takes place around $\kform$. As mentioned previously, for momenta $\kform^2 < \k^2 < \Qmed^2$ the momentum broadening of the ``in-in" contribution sets it apart from the out-of-medium contribution, lower row of eq.~(\ref{eq:r-out-of-medium}), preventing the appearance of a pure vacuum component.\footnote{There is, of course a non-zero probability that the gluon does not experience further broadening while traversing the medium and reaches the final-state cut with momentum $\k = \k'$. This contribution is indeed the genuine vacuum contribution.} For $\k^2 > \Qmed^2$ these contributions cancel exactly with part of the ``in-in" contribution, as anticipated in the discussion below eq.~(\ref{eq:r-in-in-short-leading}).

\subsection{Leading behavior of medium-induced radiation}
\label{sec:indep-leading}

In summary, the single-gluon spectrum off an accelerated charge in the presence of a medium consists of three parts. First and foremost, the induced component of the independent spectrum is given by
\beq
\label{eq:r-in-in-final-leading}
\Rq^\text{med} \approx 4\omega  \int_0^L dt'  \int\frac{d^2\k'}{(2\pi)^2}\P(\k-\k',L-t')\sin\left(\frac{\k'^2}{2 \kform^2}\right)e^{-\frac{\k'^2}{2 \kform^2}}\,.
\eeq
This is a novel, transparent way of writing the BDMPS-Z spectrum. Let us recap the main features of this spectrum. It describes the emission of a gluon with momentum $\k'$, distributed mainly around the preferred value $\kform$ which corresponds to the amount of momentum accumulated during its formation time $\tform$. After the gluon is formed it is no longer correlated with the emitting quark and its subsequent Brownian motion along its trajectory leads to a characteristic momentum broadening. This spectrum scales with the length of the medium $L$, since the medium-induced emissions can take place at any point along the trajectory of the quark through the medium. The remaining terms are not enhanced neither by the medium length $L$ nor are they enhanced by a logarithmic divergence, such as for the bremsstrahlung, and they can therefore be neglected at the level of our approximations.

In addition, one has the soft and hard bremsstrahlung contributions which are described in detail in sec.~\ref{sec:independent-inout}.

\subsection{Analytical continuation prescription for short formation times}
\label{sec:indep-revisited}

While the independent spectrum by itself permits a fully analytical discussion, see the previous section, this is not the case for the interferences. To highlight the interesting features of these contributions in a well-controlled manner we will therefore introduce a procedure which captures the leading behavior of the independent spectrum around the typical medium scale at emission, $\k'^2 \sim \kform^2$, and which still permits an analytical treatment. In the subsequent sections we will show that this procedure also can be applied to the interference spectra.

\begin{figure}
\centering
\includegraphics[width=0.35\textwidth]{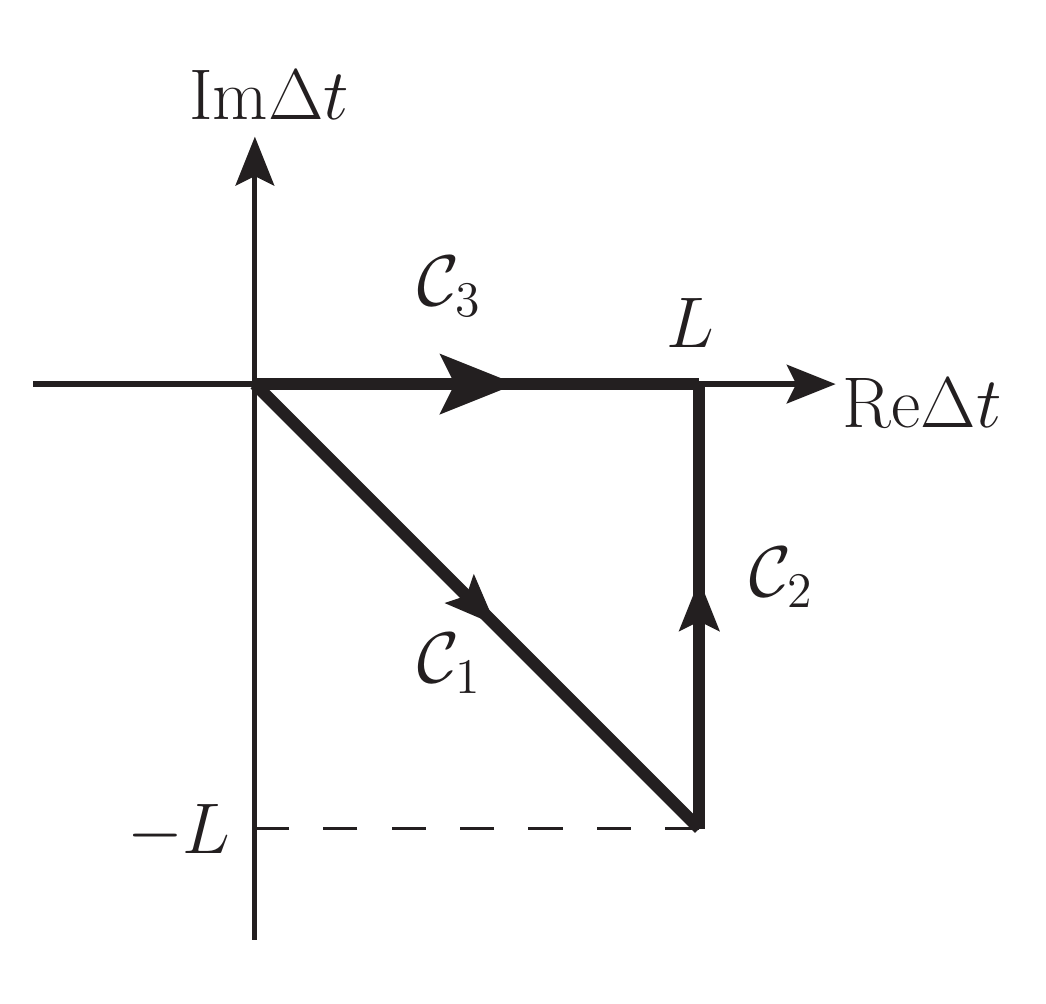}
\caption{The analytic continuation for the $\Delta t$ integration.}
\label{fig:contour}
\end{figure}

The main lesson learned from the considerations in section~\ref{sec:independent-analysis} is that the time difference of emission in the amplitude and complex conjugate amplitude, denoted by $\Delta t$, is limited by the formation time, $\tform$, due to the LPM suppression. This is, e.g., clearly seen in the factor $\cos^{-2}(\Omega \Delta t)$ in eq.~(\ref{eq:r-in-in-1}) which causes an exponential suppression of the spectrum at large $\Delta t$, since
\beq
\label{eq:cos-denominator}
\cos \left( \Omega \Delta t \right) \sim e^{\Delta t/\tform}, \quad \text{for } \Delta t \gg \tform \,.
\eeq
These considerations indicate that  the leading behavior of the spectrum is generated in the phase space where $\Delta t < \tform$. Exploiting this fact, let us rewrite the limits of integration of eq.~(\ref{sec:independent-inin}) as follows
\beq
\int_0^Ldt' \int_0^{t'} dt = \int_0^L d\Delta t \int_{\Delta t}^L dt' \,.
\eeq
Following the logic of sec.~\ref{sec:independent-inin}, we divide the latter integration over $t'$ into two pieces at the formation time.\footnote{At the level of this approximation we do not control the numerical factors related to this arbitrary separation, cf. the parameter $c$ from sec.~\ref{sec:independent-inin-early}.} Hence, for the upper part of the $t'$ integral, i.e. when $\tform < t' < L$, the two integrals decouple and eq.~(\ref{eq:r-in-in-1}) reads, in this case
\begin{align}
\left. \Rq^\text{in-in} \right|_{t' > \tform} =& \; 2 \text{Re} \int_{\tform}^{L} dt' \int\frac{d^2\k'}{(2\pi)^2} \mathcal{P}(\k-\k',L-t') \int_0^L d \Delta t  \frac{\k'^2}{\cos^2\left(\Omega \Delta t \right)} \nn
& \quad \times\exp \left[- i \frac{\k'^2}{2 \omega \Omega}\tan \left( \Omega \Delta t\right) \right] \,.
\end{align}
Instead of proceeding to calculate the integral exactly, as done in the previous section, we follow an alternative route which captures the main features of the spectrum. Performing an analytic continuation to the complex plane of $\Delta t$, see fig. \ref{fig:contour}, we note that the sum of the integrals along the three contours $\mathcal{C}_1$, $\mathcal{C}_2$ and $-\mathcal{C}_3$ should give zero since there are no poles inside the integration contour. Then, the contour along the real axis, $\mathcal{C}_3$ in fig. \ref{fig:contour}, is equal to the sum of integrals along the trajectories, which are defined as
\begin{enumerate}
\item $\mathcal{C}_1$: $\Delta t = (1-i)x$ for $x \in [0,L]$,
\item $\mathcal{C}_2$: $\Delta t = L(1 + i x)$ for $x \in [-1 ,0]$.
\end{enumerate}
Note that for the second integration contour, the dependence on $\Delta t$ always will scale like $L /\tform \gg 1$. In other words, the integrand along this contour is always exponentially suppressed and we can neglect this contribution altogether. Therefore, the total integral becomes equal to the integral along contour $\mathcal{C}_1$ and reads
\begin{align}
\left. \Rq^\text{in-in} \right|_{t' > \tform} \simeq &\; 2 \text{Re} \int_{\tform}^{L} dt' \int\frac{d^2\k'}{(2\pi)^2} \mathcal{P}(\k-\k',L-t') \int_0^L d x  \frac{(1-i)\k'^2}{\cosh^2\left(x /\tform \right)} \nn
& \quad \times\exp \left[-  \frac{\k'^2}{2 \omega \Omega}\tanh \left( x/\tform\right) \right] \,.
\end{align}
The leading contribution to this integral comes from the region of small-$x$, $x < \tform$, where we can expand the hyperbolic tangent in the integrand. Introducing $\tform$ as the effective cut-off of this integral, we get
\begin{multline}
\left. \Rq^\text{in-in} \right|_{t' > \tform} \simeq 2 \text{Re} \int_{\tform}^{L} dt' \int\frac{d^2\k'}{(2\pi)^2} \mathcal{P}(\k-\k',L-t') \int_0^{\tform}d x (1-i)\k'^2 \exp \left[-  (1+i) \frac{\k'^2}{2 \omega}x \right] \,.
\end{multline}
Finally, after performing the remaining integration over $x$, we get
\beq
\label{eq:r-in-in-approx-1}
\left. \Rq^\text{in-in} \right|_{t' > \tform} \simeq 2 \text{Re} \int_{\tform}^{L} dt' \int\frac{d^2\k'}{(2\pi)^2} \mathcal{P}(\k-\k',L-t') \sin \left( \frac{\k'^2}{2 \kform^2} \right) \exp \left(- \frac{\k'^2}{2 \kform^2} \right) \,,
\eeq
which forms part of the genuine, medium-induced spectrum. We will recover the missing piece, i.e., for $t' < \tform$, from the second integration region.

Now, assume that $\Delta t< t' < \tform$. We are interested in the regime where $\tform \ll L$, as usual, so we can safely approximate $\mathcal{P}(\k-\k',L-t') \approx \mathcal{P}(\k-\k',L)$. After performing the now trivial integration over $t'$, we can rewrite the spectrum as
\begin{align}
\left. \Rq^\text{in-in}\right|_{t' < \tform} = & \; 2 \text{Re} \int\frac{d^2\k'}{(2\pi)^2} \mathcal{P}(\k-\k',L)  \int_0^L d\Delta t \frac{\k'^2}{\cos^2\left(\Omega \Delta t \right)} \nn 
& \quad \times \left. \frac{\partial}{\partial \alpha}\exp \left[- i \frac{\k'^2}{2 \omega \Omega}\tan \left( \Omega \Delta t\right) + (\tform - \Delta t) \alpha \right] \right|_{\alpha = 0} \,.
\end{align}
Introducing the same continuation prescription as for the previous integration region, we get
\begin{align}
\left. \Rq^\text{in-in}\right|_{t' < \tform} \simeq& \; 2 \text{Re} \int\frac{d^2\k'}{(2\pi)^2} \mathcal{P}(\k-\k',L)  \int_0^{\tform} dx (1-i) \k'^2 \nn
&\quad \times \left. \frac{\partial}{\partial \alpha}\exp \left\{-\left[ (1+ i) \frac{\k'^2}{2 \omega}x + (1-i) \alpha \right] x + \tform \alpha \right\} \right|_{\alpha = 0} \,,
\end{align}
and after some algebra, we find
\begin{align}
\label{eq:r-in-in-approx-2}
\left. \Rq^\text{in-in}\right|_{t' < \tform} \simeq &\, 2 \text{Re} \int\frac{d^2\k'}{(2\pi)^2} \mathcal{P}(\k-\k',L)  \Bigg[ \frac{4\omega^2}{\k'^2} \left(1- e^{-(1+i)\frac{\k'^2}{2\kform^2}} \right) \nn 
& \quad- 2i \omega \tform \left(1 - e^{-(1+i) \frac{\k'^2}{2 \kform^2}} \right) - 2(1+i) \omega \tform \, e^{-(1+i) \frac{\k'^2}{2 \kform^2}} \Bigg] \,.
\end{align}
The first line of eq.~(\ref{eq:r-in-in-approx-2}) contains the hard bremsstrahlung term, cf. eq.~\eqref{eq:r-in-in-short-leading} which is suppressed for $\k'^2 < \kform^2$, as expected. The second line, on the other hand, contains the missing piece of the independent spectrum in eq.~(\ref{eq:r-in-in-approx-1}), i.e., that scales like $\tform$ whereas eq.~(\ref{eq:r-in-in-approx-1}) is proportional to $L-\tform$. The remainder in eq.~(\ref{eq:r-in-in-approx-2}) scales also as $\tform$ and will be neglected at the level of the present approximation. 

In consequence we have recovered the correct leading behavior of the medium-independent spectrum, cf. section~\ref{sec:indep-leading}. Introducing an analytic continuation of the integration was necessary to correctly treat the complex trigonometric functions in the integrand. This trick also allow us to obtain the correct exponential cut-off of the emission spectrum at $\k'^2 > \kform^2$. Thus, keeping only the leading part of the spectrum for $\tform \ll L$ we were able to find the main characteristics.

\section{The interferences --- short formation times}
\label{sec:interferences}

Continuing now to the main discussion about the interferences in the presence of a medium, we remind the reader that we will mainly focus on the region of large-angle radiation, i.e. radiation at angles $\theta \gg \theta_\qqb$, which automatically implies that $|\delta \k| \ll r_\perp^{-1}$, $\Qmed$. In any case, at angles $\theta \ll \theta_\qqb$ the interferences are suppressed and the spectrum becomes the superposition of the two independent components. 

The interference contribution in eq.~(\ref{eq:j-inin-1}) has a much more complicated structure than the independent one, cf. eq.~(\ref{eq:r-in-in-1}). Yet the arguments presented in the previous sections, most importantly about the dominance of short formation times, allow us to simplify the expressions and extract analytically the leading behavior of the spectrum. Note first that all combinations of $A$'s and $B$'s in eq.~(\ref{eq:j-inin-1}) can be written in terms of the cosine and tangent trigonometric functions. Since we only will consider the region $\Delta t \lesssim \tform$ we can safely approximate all $\cos (\Omega \Delta t) \approx 1$ --- in the opposite case, we will treat the spectrum as exponentially suppressed, cf. eq.~(\ref{eq:cos-denominator}). Then, the ``in-in" component of the interference spectrum reads
\begin{align}
\label{eq:j-in-in-2}
\J^\text{in-in} \simeq&  \;\text{Re}  \int_0^L dt \int_0^{L-t} d\Delta t \int\frac{d^2\k'}{(2\pi)^2} \, \big[ 1 - \Delta_\text{med}(t) \big] \mathcal{P} (\bkappa - \k', L- t - \Delta t ) \nn
&\quad \times \k' \cdot \left[\bar \k' - \Omega t \tan \left( \Omega \Delta t\right) \delta \k\right] \nn
&\quad \times \exp \left[- i \frac{(\k' + \bar \k' ) \cdot \pip}{2} t  -i \left(\k'^2 - \Omega^2 t^2 \delta \k^2 \right) \frac{\tan (\Omega \Delta t)}{2 \omega \Omega} \right] \,+\, \text{sym.} \,,
\end{align}
where we also have reorganized the integration limits. Recall that $\delta \k$, appearing explicitly in eq.~(\ref{eq:j-in-in-2}), is the scale that restores coherence in vacuum, see eq.~(\ref{eq:rsing-vac}). Within the same scheme, the ``in-out" contribution reads
\begin{align}
\label{eq:j-in-out-2}
\J^\text{in-out} \simeq & \; \text{Re}  \int_0^L dt \, \big[ 1 - \Delta_\text{med}(t) \big]  \frac{2i\omega}{\cos^2[ \Omega(L-t)] \, \bkappa^2} \bkappa \cdot \left\{ \bbkappa - \Omega t \tan \left[ \Omega (L-t)\right] \delta \k \right\} \nn
&\quad \times \exp \left[- i \frac{(\bkappa + \bbkappa ) \cdot \pip}{2} t  -i \left(\bkappa^2 - \Omega^2 t^2 \delta \k^2 \right) \frac{\tan [\Omega (L-t)] }{2 \omega \Omega} \right] \,+\, \text{sym.} 
\end{align}
and, finally, the ``out-out" component is given by eq.~(\ref{eq:j-outout-1}). Note, that since we only have one time-integration in eq.~(\ref{eq:j-in-out-2}) we can solve this spectrum exactly within our approximation and have therefore kept the cosine in the denominator of eq.~(\ref{eq:j-in-out-2}). Since we are assuming that the scales induced by the medium are larger than the vacuum one, we will show below, separately for the ``dipole" and ``decoherence" regimes, when terms involving the recurring combination of $\Omega t\, \delta \k$ can be neglected. Along the same lines, we will presently only focus on radiation that takes place outside of the cone of the $\qqb$-pair, i.e., for $\theta \gg \theta_\qqb$. Formally, we take the limit of small opening angles $\pip \to 0$ only after including the symmetrical pieces in eqs.~(\ref{eq:j-in-in-2}) and (\ref{eq:j-in-out-2}) in proper way. This condition implies the following constraint on the relevant timescales
\beq
\label{eq:time-inside-cone}
\tform < \tdip \,,
\eeq
which implies that $\theta_f > \theta_\qqb$, where $\theta_f$ is the characteristic angle of medium-induced emissions at formation time. These approximations allow us to separate the relevant time- and momentum-scales of the problem such that an analytical study becomes feasible. In the general case, only a numerical analysis can further guide our insight, see sec.~\ref{sec:numerics}.

As already pointed out in sec.~\ref{sec:discussion}, the onset of decoherence appears chiefly through the factor $(1-\Delta_\text{med}(t,0))$ which is responsible for damping out all interferences \cite{MehtarTani:2011tz}. Following closely the discussion in the preceding section we discuss separately the interference spectrum in the ``dipole" and ``decoherence" regimes, respectively.

\subsection{The ``dipole" regime}
\label{sec:interferences-dipole}

In the ``dipole" regime the color correlation of the $\qqb$-pair prior to the emission partly survives the passage through the medium. Recalling the discussion in sec.~\ref{sec:discussion-dipole}, we define this regime by the condition
\beq
L \ll \tdip \,,
\eeq
see tab.~\ref{tab:regimes} for further details. To reduce the complexity of the calculation and since the dominating scale in this regime is $\Qdip \gg \Qmed$ we neglect for the time being the effect of broadening.\footnote{The effect of the momentum broadening is mainly to smear out the $k_\perp$-distribution, which does not affect the physical picture discussed in this regime.} In this regime we expand the decoherence parameter, obtaining
\beq
\label{eq:dipole-decoh-param}
1-\Delta_\text{med}(t,0) \simeq 1 - \frac{1}{12}\hat q \theta_\qqb^2 t^3 \,.
\eeq
The former contribution, the one proportional to the numerical factor 1, is responsible for the cancellation of the vacuum singlet spectrum and the medium-induced independent spectrum in their respective regions, as will be discussed in detail below. Since we will separate the analysis of all the contributions proportional to this term, we will therefore label the corresponding expressions with the subscript `(0)' (alluding to the order of expansion of $1-\Delta_\text{med}$). The latter piece in eq.~(\ref{eq:dipole-decoh-param}), which is proportional to the square of the dipole-size, $r_\perp(t) = \theta_\qqb t$, comes with a positive contribution to the spectrum and establishes what we call ``partial decoherence," and we label it with the corresponding subscript `(1).'

At the outset, let us try to simplify eqs.~(\ref{eq:j-in-in-2}) and (\ref{eq:j-in-out-2}) even further by analyzing the conditions that have to be met to allow us to drop the terms proportional to $\delta \k$. From previous discussion, we know that $\Delta t$ should be smaller than or of the order of $\tform$ while the emission time can reach up to $L$. The latter point allow us to expand the argument of the tangents in eq.~(\ref{eq:j-in-in-2}). Apart from the decoherence parameter, the pre-factor of this equation can be written as
\beq
\k \cdot \left[\bar \k + i \frac{t\, \Delta t }{\tform^2}\delta \k \right] \,.
\eeq
As a first estimate of the corrections related to the second term in the brackets, we substitute $\Delta t$ and $t$ with their respective maximal values. Thus, this factor can be dropped as long as
\beq
\label{eq:simplify-condition-dipole}
\frac{L}{\tform} \frac{|\delta \k|}{|\k|} \sim \frac{\theta_\qqb}{\theta}\left(\frac{\thform}{\thcrit} \right)^{2/3} \ll 1\,.
\eeq
Here we have introduced the characteristic angle $\thcrit = 1/\sqrt{\hat q L^3}$, which appears for the maximal value of the decoherence parameter, i.e., when $t=L$ in eq.~(\ref{eq:decoh-parameter}), for instance. In the ``dipole" regime we automatically have $\theta_\qqb \ll \thcrit$, see tab.~\ref{tab:regimes}. Note, on the other hand, that the condition of short formation times also implies that $\thcrit \ll \thform$. For the region we are interested in we therefore have the following hierarchy of scales
\beq
\theta_\qqb \ll \thcrit \ll \thform \,,
\eeq
see also eq.~(\ref{eq:time-inside-cone}). Therefore, as long as the gluon emission angle is large enough, $\theta \gtrsim \thform$, the condition in eq.~(\ref{eq:simplify-condition-dipole}) is fulfilled. This discussion also applies to the second term in the exponential eq.~(\ref{eq:j-in-in-2}), proportional to $\delta \k^2$, and for the $\mathcal{J}^\text{in-out}$ component in eq.~(\ref{eq:j-in-out-2}).

The behavior of the interferences in the ``dipole" regime is quite intricate since it involves all components which should be directly compared both to the vacuum spectrum and the independent medium-induced one. The main features that we derive below are therefore also conveniently summarized in sec.~(\ref{sec:interferences-dipole-summary}).

\subsubsection{The ``in-in" component}

Since the hardest scale of the problem is $\Qdip \gg \Qmed$ we will neglect the effect of broadening in this regime, replacing $\mathcal{P}(\k-\k',L-t) = (2\pi)^2\delta(\k-\k')$. Since this process does not contribute to the final-state yield of produced gluons we do not expect this simplification to change the physical picture of the emission process itself. See also appendix~\ref{sec:broadening}. This simplification makes feasible an analytic discussion of the various components of the interference spectrum. We present numerical results including all effects of the full spectrum in sec.~\ref{sec:numerics} and postpone a detailed assessment of the approximations employed below to that section. 

Let us presently investigate the ``in-in" component at large angles, and as a fist step only the part which comes with the unity in eq.~(\ref{eq:dipole-decoh-param}). This contribution reads
\begin{align}
\label{eq:j-in-in-dipole-0}
\mathcal{J}^\text{in-in}_{(0)} &\simeq \text{Re} \int_0^Ld\Delta t \int_{\Delta t}^L dt' \bkappa \cdot \bbkappa \exp \left[-i \frac{(\bkappa+\bbkappa) \cdot \pip}{2}(t' - \Delta t) -i \frac{\bkappa^2}{2 \omega \Omega} \tan \Omega \Delta t \right] \,+\, \text{sym.} \nn
&\simeq \frac{4 \omega^2 \bkappa \cdot \bbkappa}{\bbkappa^2 - \bkappa^2} \,\text{Re} \left\{\frac{\mathcal{F}(\bkappa)}{\bkappa^2} \left[1 + e^{i \frac{(\bkappa + \bbkappa) \cdot \r}{2}} \right] - \frac{\mathcal{F}(\bbkappa)}{\bbkappa^2} \left[1 + e^{-i \frac{(\bkappa + \bbkappa) \cdot \r}{2}} \right]\right\} \,.
\end{align}
When going to the second line in eq.~\eqref{eq:j-in-in-dipole-0}, we have applied the procedure of analytical continuation described in sec.~\ref{sec:indep-revisited}, but note that we have not performed the $\pip \to 0$ limit yet. Furthermore, in eq.~(\ref{eq:j-in-in-dipole-0}) we have defined the function
\beq
\mathcal{ F}(\k)  = 1 - \exp \left[-(1+i)\frac{\k^2}{2\kform^2} \right]\,.
\eeq
Consider the situation when $\kform \sim k_\perp \ll \Qdip$. Expanding the relevant phases to first order and combining the symmetrical terms, we find
\beq
\label{eq:j-in-in-dipole-1}
\mathcal{J}^\text{in-in}_{(0)} \simeq 4\omega L \sin \left[ \frac{\k^2}{2 \kform^2} \right] \exp \left[ -\frac{\k^2}{2 \kform^2}\right] + \frac{8 \omega^2}{\k^2} \left\{1 - \cos \left[ \frac{\k^2}{2 \kform^2} \right] \exp \left[ -\frac{\k^2}{2 \kform^2}\right]\right\}  \,.
\eeq
We have thus recovered the medium-induced independent spectrum in the first term of eq.~(\ref{eq:j-in-in-dipole-1}), which will cancel with the medium-induced component $\mathcal{R}_q^\text{in-in}$ in this regime. This spectrum also has the same behavior as $\mathcal{R}_q^\text{in-in}$ in the hard sector. In particular, for $\kform \ll k_\perp$ we obtain, directly from eq.~(\ref{eq:j-in-in-dipole-0}),
\beq
\label{eq:j-in-in-dipole-11}
\left. \mathcal{J}^\text{in-in}_{(0)} \right|_{k_\perp \gg \kform} \simeq \frac{4 \omega^2}{\k^2} \big[1 + \cos \left( \k \cdot \r\right) \big] \,,
\eeq
which is a vacuum-like contribution. In particular, in the interval $\kform \ll k_\perp \ll \Qdip$ we recover a vacuum contribution, cf. eq.~(\ref{eq:r-in-in-short-leading}), which should be cancelled by the corresponding contribution in $\mathcal{J}^\text{in-out}_{(0)}$. It is also interesting to comment the oscillatory behavior of the spectrum for $k_\perp \sim \Qdip$, which already turned up in $\mathcal{J}^\text{out-out}$, see eq.~(\ref{eq:j-outout-1}). Such behavior is an artifact due to the sharp boundary condition at the medium border, i.e. at $L$. In fact, as we demonstrate below, this oscillatory behavior cancels out exactly at $k_\perp \gg \Qdip$. Then, quite astonishingly, what remains is a genuine, physical vacuum contribution which comes into play at large transverse momenta.

Next, we analyze the piece which is proportional to $r_\perp^2$, i.e., the one multiplying the second term of eq.~(\ref{eq:dipole-decoh-param}). It reads
\begin{align}
\label{eq:j-in-in-dipole-2}
\mathcal{J}^\text{in-in}_{(1)} \simeq & \; -\text{Re} \int_0^L d\Delta t \int_{0}^{L-\Delta t} dt \, \Delta_\text{med}(t) \, \bkappa \cdot \bbkappa \nn 
& \quad \times \exp \left[-i \frac{(\bkappa+\bbkappa) \cdot \pip}{2}t -i \frac{\bkappa^2}{2 \omega \Omega} \tan \Omega \Delta t \right] \,+\, \text{sym.} \,,
\end{align}
where explicitly $\Delta_\text{med}(t) = \hat q \theta_\qqb^2 t^3/12$. In the general case, the integration over $t$ results in a quite tedious expression which simplifies considerably when taking the appropriate limits with respect to the dipole scale $\Qdip$. Starting from small momenta, when $k_\perp \ll \Qdip$, we find
\beq
\int_0^{L-\Delta t} \!dt \,\Delta_\text{med} (t) \exp \left[-i \frac{(\bkappa+\bbkappa) \cdot \pip}{2}t \right] \simeq \frac{(L- \Delta t) \,\Delta_\text{med}(L-\Delta t)}{4} \,,
\eeq
plus terms that are at least suppressed like $\mathcal{O}(\k \cdot \r)$. Furthermore, knowing that $\Delta t$ will be limited by $\tform$, and thus much smaller than $L$, we keep only the linear term in $\Delta t$, i.e., $(L-\Delta t) \,\Delta_\text{med}(L-\Delta t) \simeq (L - 4 \Delta t)\, \Delta_\text{med}(L)$. Thus, in this domain the contribution becomes
\beq
\label{eq:j-in-in-dipole-3}
\left. \mathcal{J}^\text{in-in}_{(1)} \right|_{k_\perp \ll \Qdip} \simeq - \Delta_\text{med}(L) \text{Re} \left\{ -i \omega L \mathcal{F}(\k) + \frac{8 \omega^2}{\k^2} \mathcal{F}(\k)- (1+i) 4 \omega \tform \big[1 - \mathcal{F}(\k) \big] \right\} \,.
\eeq
The first component in eq.~(\ref{eq:j-in-in-dipole-3}) is equivalent to the independent spectrum multiplied by a factor $-\Delta_\text{med}/4$ and in the second component we once again recover the vacuum-like spectrum from very early emissions. The latter factor in eq.~(\ref{eq:j-in-in-dipole-3}) is sub-leading, $\propto \tform$, and will be neglected in the following. 

Proceeding to the second situation, the integral over $t$ in eq.~(\ref{eq:j-in-in-dipole-2}) in the domain where $k_\perp \gtrsim \Qdip$ gives
\begin{align}
\int_0^{L- \Delta t} \!dt \,\Delta_\text{med} (t) \exp \left[-i \frac{(\bkappa+\bbkappa) \cdot \pip}{2}t \right] & = \frac{-2i \omega}{\bbkappa^2 - \bkappa^2} \Delta_\text{med}(L -\Delta t) \nn
& \quad \times \exp\left[ -i \frac{(\bkappa + \bbkappa) \cdot \pip}{2} (L- \Delta t) \right] \,,
\end{align}
plus terms that are at least suppressed like $\mathcal{O}[(\k \cdot \r)^{-1}]$. In this case, the decoherence parameter is approximated simply by its leading contribution, i.e., $\Delta_\text{med}(L-\Delta t,0) \simeq \Delta_\text{med}(L,0)$. Since in this case we are already above the medium scale, since $\kform \ll \Qdip$, we can drop any exponentially suppressed term to get
\beq
\left. \mathcal{J}^\text{in-in}_{(1)} \right|_{k_\perp \gtrsim \Qdip} \simeq - \Delta_\text{med}(L) \frac{4 \omega^2}{\k^2}  \cos \left( \k \cdot \r \right) \,.
\eeq
This concludes our investigation of the ``in-in" component in the ``dipole" regime.

\subsubsection{The ``in-out" component}

From the analysis of the preceding sections, we see that only emissions roughly one formation time away from the medium border will contribute to the ``in-out" spectrum. We will also only consider large-angle emissions where, following the discussion in sec.~(\ref{sec:interferences-dipole}), we can drop the terms proportional to $\delta \k$ in eq.~(\ref{eq:j-in-out-2}). After rearranging the integral limits, the dominant contribution reads
\begin{align}
\J^\text{in-out} \simeq &\; - \text{Re} \int_0^{L} dt \big(1 - \Delta_\text{med}(L-t) \big)\frac{2i\omega\, \bkappa \cdot \bbkappa}{\bkappa^2 \, \cos^2( \Omega t)} \nn
&\quad \times \exp\left[-i \frac{(\bkappa + \bbkappa) \cdot \pip}{2}(L-t)  -i\frac{\bkappa^2}{2\omega \Omega} \tan(\Omega t) \right] \,+\, \text{sym.}
\end{align}
Since the integral over $t$ will be cut off above $\tform$ we can neglect the subleading finite-size corrections to the argument of the decoherence parameter, $1-\Delta_\text{med}(L-t,0) \approx 1-\Delta_\text{med}(L,0)$. Then, the remaining integral domain yields
\beq
\J^\text{in-out} \simeq - \big(1 - \Delta_\text{med}(L) \big) \frac{4\omega^2 \, \bkappa \cdot \bbkappa}{\bkappa^2 \bbkappa^2} \, \text{Re} \,\mathcal{F}(\bbkappa) \exp\left[-i \frac{(\bkappa + \bbkappa) \cdot \r}{2}\right] \,+\, \text{sym.}
\eeq
where the $\pip \to 0$ limit is not yet taken. As for the independent ``in-out" component, this spectrum is suppressed for $k_\perp < \kform$. Above the medium scale, $\kform \ll k_\perp$, $\text{Re} \,\mathcal{F}(\k) \simeq 1$ and the spectrum simply reads
\beq
\label{eq:j-in-out-dipole-1}
\left.\J^\text{in-out} \right|_{k_\perp \gg \kform} \simeq - \big(1 - \Delta_\text{med}(L) \big) \frac{8\omega^2}{k^2} \cos\left( \k\cdot \r \right) \,.
\eeq
As previously anticipated, this term will fully cancel the spurious vacuum-like contribution appearing in $\mathcal{J}^\text{in-in}$ for emissions taking very place very early on.

\subsubsection{Summary of the ``dipole" regime}
\label{sec:interferences-dipole-summary}

To have all relevant expressions in one place, let us also remind the reader of the ``out-out" component, which at large angles is given by
\beq
\label{eq:j-out-out-2}
\J^\text{out-out} = \big[1 - \Delta_\text{med}(L) \big] \frac{4\omega^2}{\k^2}  \cos\left( \k\cdot \r \right) \,.
\eeq
Recall also that the medium-induced spectrum $\mathcal{R}_q^\text{in-in} + \mathcal{R}_q^\text{in-out}$ only has support for $k_\perp < \kform$, see eq.~(\ref{eq:r-in-in-final-leading}), while the vacuum contribution $\mathcal{R}^\text{out-out}_q$ extends up to arbitrary $k_\perp$. Summarizing the results for the ``dipole" regime obtained in this section, we will highlight the main features in the different kinematical regimes that we assume, as always, to be clearly separated. 

The singlet spectrum is decomposed into a quark and an antiquark coherent part, $\mathcal{R}_\text{sing} = \mathcal{P}_q + \mathcal{P}_{\bar q}$, where $\mathcal{P}_q = \mathcal{R}_q - \mathcal{J}_q$, and, as usual, we only focus on the quark part in the following. Below, both the independent and interference components are a sum of their vacuum and medium contributions, respectively. 

\begin{description}
\item[$\mathbf{k_\perp < \delta k_\perp}$:] This is the regime where interferences are always suppressed. The only contribution to the spectrum arises from the logarithmically divergent vacuum spectrum.

\item[$\mathbf{\delta k_\perp \ll k_\perp \lesssim \kform}$:] In this regime the independent spectra are fully cancelled by the interferences. What remains are terms proportional to the decoherence parameter; firstly, a vacuum-like contribution coming from $\mathcal{J}_{q,\, (1)}^\text{out-out}$ which reflects ``partial decoherence." Also, somehow unexpectedly, there is an additional surviving component in $\mathcal{J}^\text{in-in}_{q,\,(1)}$ which behaves as the independent component but is reduced by a factor $\Delta_\text{med}(L)$ compared to the former. The coherent spectrum off the quark is then given by
\beq
\label{eq:pqsummary-dipole-1}
\mathcal{P}_q \big|_{\delta k_\perp \ll k_\perp \lesssim \kform} = \Delta_\text{med}(L) \left[ \frac{4\omega^2}{ \k^2} + \omega L \sin\left( \frac{\k^2}{2 \kform^2} \right) \exp \left( \frac{\k^2}{2 \kform^2} \right)\right] \,.
\eeq

\item[$\mathbf{\kform \ll k_\perp \ll \Qdip}$:] As in the previous domain, a vacuum-like behavior is obtained. This demonstrates how ``partial decoherence" extends above the induced medium scale. Note therefore that in the ``dipole" regime we obtain a logarithmic enhancement or, in other words, a hardening of the spectrum. This was already noted in \cite{MehtarTani:2011gf}. The coherent spectrum is simply given by
\beq
\label{eq:pqsummary-dipole-2}
\mathcal{P}_q \big|_{\kform \ll k_\perp \ll \Qdip} = \Delta_\text{med}(L) \frac{4\omega^2}{ \k^2} \,.
\eeq
With the inclusion of broadening, which we so far have neglected in the discussion, we recover a pure vacuum contribution when the medium scale $\kform$ is substituted with $\Qmed$, see also the discussion in secs.~\ref{sec:independent-inin-early} and \ref{sec:independent-inout}.

\item[$\mathbf{ \Qdip \lesssim k_\perp}$:] A `novel' hard component is generated from early emissions in $\mathcal{J}^\text{in-in}_{(0)}$, i.e. proportional to the factor 1 in eq.~(\ref{eq:j-in-in-dipole-11}), which is the only piece that remains after the cancellation between all the interference components. In particular, the genuine ``out-out" interference contribution is completely cancelled. Thus, the remaining vacuum-like piece, which does not contain information about the medium or the dipole, cancels $\mathcal{R}_q^\text{out-out}$ in this domain. This establishes the onset of coherence in the ``dipole" regime, leaving us with a strongly suppressed coherent spectrum at transverse momenta larger than the hardest scale of the problem,
\beq
\label{eq:pqsummary-dipole-3}
\mathcal{P}_q \big|_{\Qdip \lesssim k_\perp} \simeq 0\,.
\eeq
\end{description}
We have described the features of ``partial decoherence" and the restoration of coherence above the the hard scale in the ``dipole" regime in rough accordance with the general characteristics outlined in sec.~\ref{sec:discussion-dipole}. Nevertheless, due to the highly complex dynamics of this regime, note e.g. the unforeseen `independent'-like contribution to $\mathcal{J}^\text{in-in}_{(1)}$ in eq.~(\ref{eq:j-in-in-dipole-3}) which modifies the behavior close to the peak of the independent distribution, and the fact that the clear separation of scales we have assumed throughout the analytical discussion is unrealistic, the features obtained above should be taken as a first order idealization of the physics involved. The region $k_\perp \sim \kform$ is in truth badly controlled due to the additional $\delta \k$-terms that we have neglected to make the analysis feasible. In fact, numerical data in the ``dipole" regime on the double-differential angular spectrum shows a highly oscillatory and complex behavior at large angles in general and around the peak of the independent distribution in particular. Nevertheless, the dominant features outlined above emerge evidently in the energy spectrum, see sec.~\ref{sec:numerics}, where we recover an almost ideal scaling with the hard scale of the problem, $\Qdip$, in perfect agreement with the concept of the restoration of coherence.

\subsection{The ``decoherence" regime}
\label{sec:interferences-dense}

The ``decoherence" regime is characterized by the rapid color decorrelation of the $\qqb$-pair as it traverses the medium and the hard scale is given by the medium characteristics $\Qmed \gg \Qdip$. In terms of the survival probabilities discussed extensively above, this translates into a short decoherence time compared to the size of the medium, i.e. $\tdip \ll L$. Due to the presence of the decoherence parameter, see eq.~(\ref{eq:j-in-in-2}), the interferences will therefore only contribute at time-scales $t < \tdip$. In this particular domain we can safely approximate
\beq
\left. \big[ 1- \Delta_\text{med}(t) \big] \right|_{t < \tdip} \approx 1 \,.
\eeq
However, we underline that this domain of times is relevant exclusively for the ``in-in" component. The ``in-out" component has support mainly close to the medium border --- roughly one formation length in, see eq.~(\ref{eq:j-in-out-2}). Since we are interested in large-angle radiation, see eq.~(\ref{eq:time-inside-cone}), we can safely neglect it completely. The ``out-out" component, on the other hand, is simply proportional to $1 - \Delta_\text{med}(L) \approx 0$ and drops out, too. The gluon spectrum in this regime was also studied in \cite{CasalderreySolana:2011rz}.

The limiting condition on the emission time simplifies the situation considerably compared, e.g., to the situation in the ``dipole" regime. For the ``in-in" component, it implies that the integrals over $t$ and $\Delta t$ decouple completely, see eq.~(\ref{eq:time-inside-cone}). This allows us to freely perform the prescription regarding the analytic continuation of $\Delta t$ and the related approximations, see sec.~\ref{sec:indep-revisited}. As a preparatory step, let us once more estimate the magnitude of the terms multiplying $\delta \k$ in eq.~(\ref{eq:j-in-in-2}). To obtain a first estimate, let us substitute $t$ by its maximal allowed value in the ``decoherence" regime, which is just $\tdip$. Then the second term in the pre-factor becomes of the order of
\beq
\frac{\tdip}{\tform} \frac{|\delta \k|}{|\k|} \sim \frac{\big( \theta_f^2 \theta_\qqb \big)^{1/3}}{\theta}\,.
\eeq
This becomes very small if we consider large angle radiation, such that $\theta \gg \big( \theta_f^2 \theta_\qqb \big)^{1/3}$ or, roughly, 
\beq
\theta > \max \left( \theta_\qqb, \thform \right) \,.
\eeq
The additional terms will therefore contribute close to the peak of the independent distribution, $k_\perp \lesssim \kform$, but can safely neglected at the level of our approximations. Accordingly, the last factor of the exponential scales as the square of the estimate above and can therefore also be dropped.

After introducing the analytic continuation for $\Delta t$ and taking into account the simplifications discussed above, the ``in-in" spectrum reads
\begin{align}
\J^\text{in-in} & \simeq \text{Re}  \int_0^{\tdip} dt \int_0^{\tform} dx  \,(1-i)\int\frac{d^2\k'}{(2\pi)^2} \, \mathcal{P} (\bkappa - \k', L)\nn 
&\quad \times \k' \cdot \bar \k' \exp \left\{- i \frac{(\k' + \bar \k' ) \cdot \pip}{2} t  - \frac{(1+i) \k'^2}{2\omega}x \right\} \,+\, \text{sym.} \,,
\end{align}
where we have neglected the dependence of $\Delta t$ and $t$ in the broadening since $\tform, \tdip \ll L$. After performing the integral over $x$, we have
\beq
\label{eq:j-inin-dense-4}
\mathcal{J}^\text{in-in} = -\text{Re} \int_0^{\tdip} dt \int\frac{d^2\k'}{(2\pi)^2} \mathcal{P} (\bkappa - \k', L) \frac{2i\omega \, \k' \cdot \bar \k'}{\k'^2} \mathcal{F}(k') e^{- i \frac{(\k' + \bar \k') \cdot \pip}{2}t} + \text{sym.}
\eeq
where we still keep the distinction between $\k'$ and $\bar \k'$. The last phase in eq.~(\ref{eq:j-inin-dense-4}) defines a novel characteristic time-scale of the dense regime, which we denote by $\delta t$ and which reads
\beq
\delta t \sim \frac{\omega}{\bkappa^2 - \bbkappa^2} \,,
\eeq
in the absence of broadening. To get a feeling for the role of this timescale let us again compare it to the maximal available value of $t$. The ratio of the two defines a new characteristic momentum scale, $ \tdip / \delta t = k_\perp /\kcoh$, where we have defined
\beq
\label{eq:kcoh}
\kcoh \equiv \left( \hat q /\theta_\qqb \right)^{1/3} \,,
\eeq
and, correspondingly, a characteristic angle $\thcoh = \kcoh /\omega = (\thform^4/ \theta_\qqb)^{1/3}$. In the present region of interest, i.e. $\thform \ll \theta_\qqb$, it follows that $\thcoh \gg \thform$. This defines a hierarchy of scales in the dense regime, namely
\beq
\label{eq:hierarchy-scales-decoherence}
\theta_\qqb \ll \thform \ll \thcoh \,.
\eeq
Interestingly, this hierarchy is reversed when we consider radiation inside the cone. We have chosen the subscript so that to indicate that coherence is achieved for gluons with $k_\perp > \kcoh$. Note that so far in this discussion of $\delta t$ we have neglected the broadening. This mechanism will naturally provide a new hardest scale, since $\Qmed > \kcoh$ in the ``decoherence" regime, which would wash out the details of the analysis above. In the latter case, however, coherence would be restored at $k_\perp > \Qmed$, as expected from our naive picture.\footnote{We have confirmed numerically the relevance of $\kcoh$ in the absence of broadening.} Let us presently describe more carefully how this comes about.

Since we already concluded that the interferences only are operational at early times, the broadening factorizes completely from the details of the emission process, i.e., does not depend on $t$, and we can describe the latter independently of the former. Let us for that purpose define
\beq
\mathcal{J}^\text{in-in} = \int\frac{d^2\k'}{(2\pi)^2} \mathcal{P} (\bkappa - \k', L) \,\overline{\mathcal{J}^\text{in-in}} \,,
\eeq
where $\overline{\mathcal{J}^\text{in-in}}$ simply denotes the remaining terms in eq.~(\ref{eq:j-inin-dense-4}). We also point at that at large angles we can simply assume $\k' \sim \bar \k'$.
Starting with the situation when the induced gluons at emission have small transverse momenta $\k'^2 < \kcoh^2$ or, in other words, when $\tdip < \delta t$. In this case, we can neglect the last phase factor in eq.~(\ref{eq:j-inin-dense-4}) and the time integration becomes trivial, yielding
\beq
\label{eq:j-inin-dense-5}
\left. \overline{\mathcal{J}^\text{in-in}} \right|_{\k'^2 < \kcoh^2} \simeq 4\omega \tdip \sin \left(\frac{\k'^2}{2 \kform^2} \right) \exp \left(-\frac{\k'^2}{2 \kform^2} \right) \,,
\eeq
which has exactly the same structure as the independent gluon spectrum except that it scales with $\tdip$ rather than with $L$, cf. eq.~(\ref{eq:r-in-in-final-leading}), and is therefore suppressed in the ``decoherence" regime. This contribution was also identified in \cite{CasalderreySolana:2011rz}.

An important and non-negligible contribution arises in the opposite limit, however. When $\tdip > \delta t$ we have to limit the time integral by the latter, smallest time-scale. This situation applies to gluons emitted with $\k'^2 > \kcoh^2$. In this case, the ``in-in" contribution reads
\beq
\label{eq:j-inin-dense-6}
\left. \overline{\mathcal{J}^\text{in-in}} \right|_{\k'^2 > \kcoh^2} \simeq \frac{4\omega^2}{\k'^2} \,,
\eeq
and we recover a vacuum-like contribution which resembles the early emissions in the independent spectrum, cf. eq.~(\ref{eq:r-in-in-short-leading}). There is, however, a subtle difference. In contrast to the independent contribution where the vacuum-like contribution from the ``in-in" component is `regularized' by the sum of ``in-out" and ``out-out" terms (the same cancellation takes also place purely between the interference components in the ``dipole" regime), the contribution in eq.~(\ref{eq:j-inin-dense-6}) is not further reduced since $\mathcal{J}^\text{in-out} \simeq \mathcal{J}^\text{out-out} \simeq 0$ in the ``decoherence" regime. This is because the numerical factor $4$, in contrast to the factor $8$ in the independent spectrum as well as in the ``dipole" regime, corresponds exactly to the magnitude of a genuine vacuum contribution. This demonstrates that vacuum coherence is restored at large scales, i.e. above the hard scale of the medium, even in dense media. 

Keeping in mind that, due to the broadening, the dominant scale induced by the medium is in effect shifted from $\kcoh$ to $\Qmed$, cf. the discussion above, we can presently summarize our analytical results in the ``decoherence" regime.
\begin{description}
\item[$\mathbf{k_\perp < \delta k_\perp}$:] Interferences are suppressed and the only contribution to the spectrum arises from the independent components of the spectrum, see sec.~\ref{sec:independent-analysis}.

\item[$\mathbf{\delta k_\perp \gg k_\perp < \Qmed}$:] The coherent spectrum off the quark in this regime is given by
\begin{align}
\label{eq:pqsummary-decoherence-1}
\mathcal{P}_q \big|_{\delta k_\perp \ll k_\perp< \Qmed} &\simeq \frac{4\omega^2}{ \k^2} \nn
& + 4\omega (L-\tdip) \int\frac{d^2\k'}{(2\pi)^2} \mathcal{P} (\bkappa - \k', L) \sin \left(\frac{\k'^2}{2 \kform^2} \right) \exp \left(-\frac{\k'^2}{2 \kform^2} \right) \,,
\end{align}
where the first term is the vacuum bremsstrahlung. In other words, we have recovered the decoherence of the pure vacuum radiation (since the na\"ive expectation based on angular ordering would prevent vacuum radiation in this phase space) and, in addition, obtained an induced BDMPS-Z spectrum which scales as the entire medium length reduced by a small factor $(1- \tdip/L)$ due to the interferences --- recall that the ``decoherence" regime is defined by the condition $\tdip \ll L$.

\item[$\mathbf{\Qmed < k_\perp}$:] Above the hardest scale of the problem, which in this regime is $\Qmed$, we have found that the coherent spectrum is strongly suppressed,
\beq
\label{eq:pqsummary-decoherence-2}
\mathcal{P}_q \big|_{\Qmed < k_\perp} \simeq 0 \,,
\eeq
restoring coherence in this sector. This behavior is confirmed numerically in sec.~\ref{sec:numerics} and we will come back to the deeper physical meaning of this vacuum-like contribution in sec.~\ref{sec:octet}.
\end{description}

In summary, in the ``decoherence" regime all interferences, both the purely medium-induced, i.e. ``in-in" and ``in-out", components, and the vacuum-like one, encoded in the ``out-out" contribution, can be neglected up to the hard scale of the medium. In the absence of final-state momentum broadening this scale is given by $\kcoh$ in eq.~(\ref{eq:kcoh}), but in general this scale is extended up to $\Qmed$ which appears naturally in the momentum broadening factor $\mathcal{P}(\bkappa - \k',L)$. In this domain, the medium probes the two antenna constituents and induces radiation independently off each of them. Conversely, at $k_\perp > \Qmed$ (or $k_\perp > \kcoh$ if we neglect broadening, see also appendix~\ref{sec:broadening}) coherence is restored and the spectrum cancels out. This situation applies only for gluon emissions taking place at early times before the pair is completely decorrelated by the medium, i.e. at times $t < \tdip$. As long as the coherence of the pair is intact, the radiation must necessarily be coherent. Thus, for such large-angle radiation coherence (or, in other words, angular ordering) is restored due to destructive interference effects. This concludes the analytical investigations of the interferences in dense media.

\section{Numerical analysis}
\label{sec:numerics}

The most stable numerical evaluation of the gluon spectrum in the harmonic oscillator approximation is achieved in the pure coordinate-space representation. We detail the corresponding expressions in appendix~\ref{sec:HOcoordinate}. These expressions include the full effect of momentum broadening and the interplay of all the scales that we have identified in the preceding sections, and allow to study gluon production for any given kinematics. Our main strategy in this section, however, will mainly focus on putting the picture of ``decoherence" of medium-induced radiation, as argued for in secs.~\ref{sec:discussion} and \ref{sec:ho-approximation} and corroborated through a systematical, analytical investigation in secs.~\ref{sec:independent-analysis} through \ref{sec:interferences}, on a firm basis. We therefore will primarily study situations when a clear separation of dipole- and medium-scales can be acheived and search for evidence confirming the existence and role of these.

In line with our previous investigations \cite{MehtarTani:2011gf}, we choose to plot the double-differential angular spectrum of the produced gluons, $\omega dN / d\omega d\theta$, where their azimuthal angle has been integrated out. In vacuum, this spectrum would display exact angular ordering, i.e. a function $\propto \Theta(\theta_\qqb - \theta)/\theta$. Since this contribution is well known, we subtract it from the total spectrum to obtain the component explicitly depending on the presence of a medium.
\begin{figure}
\centering
\includegraphics[width=0.8\textwidth]{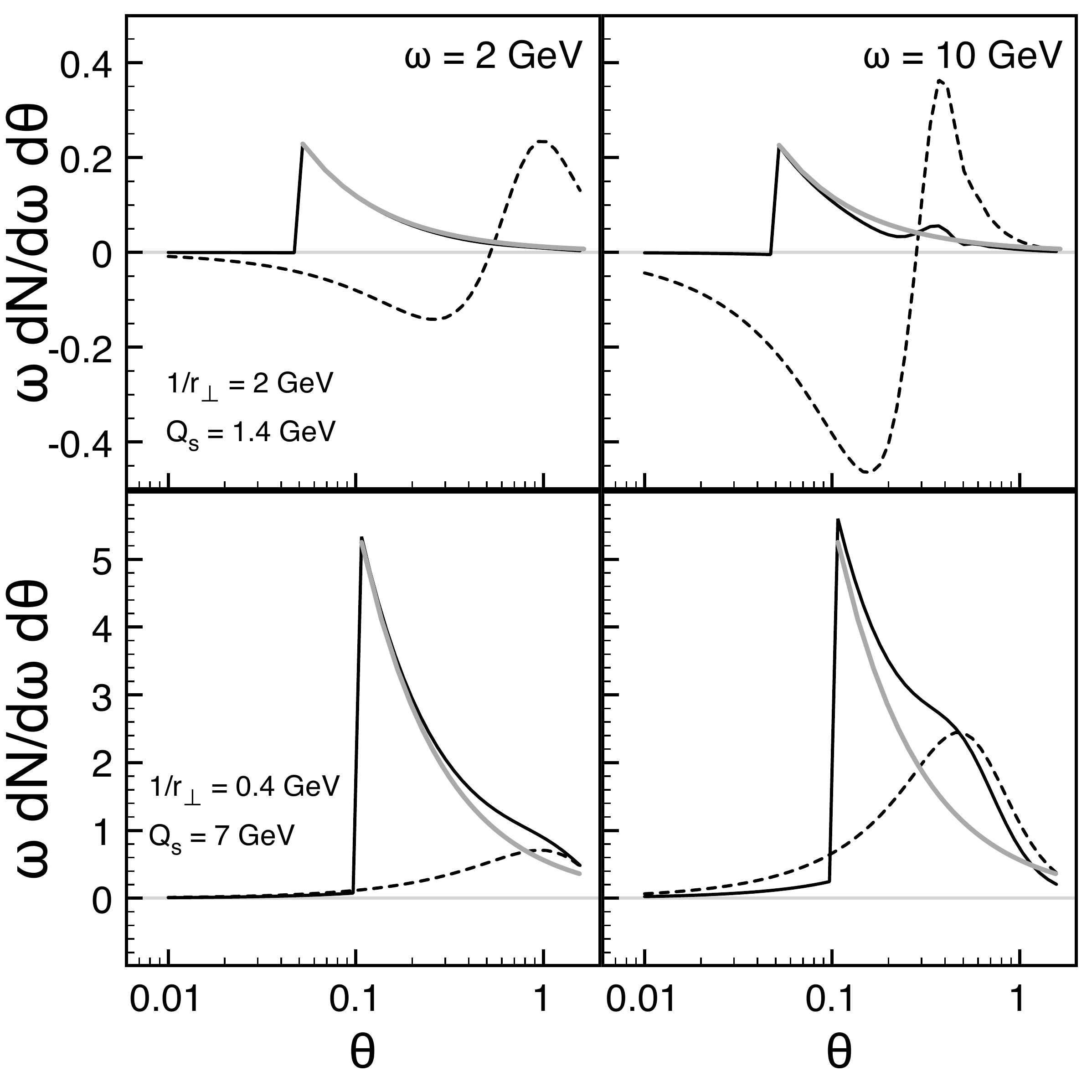}
\caption{The medium-induced gluon angular spectrum. The solid (black) line corresponds to the total coherent spectrum off the quark, cf. Eq.~\ref{eq:spectrum-coherent}, and the dashed (black) line is the independent spectrum, cf. Eq.~(\ref{eq:spectrum-independent}). To guide the eye we have also depicted the vacuum-like spectrum at large angles with a thick, grey line ($\propto \Theta(\theta - \theta_\qqb)/\theta$). See text for further details.}
\label{fig:AngularSpectrum-combined}
\end{figure}
In fig.~\ref{fig:AngularSpectrum-combined}, we plot this spectrum for a set of parameters such that the upper two panels reflect a typical situation in the ``dipole" regime (with $\Qmed = 1.4$ GeV $<$ $\Qdip = 2$ GeV), while the two lower ones depict a typical case in the ``decoherence" regime (with $\Qmed = 7$ GeV $>$ $\Qdip = 0.4$ GeV). The distributions are plotted for two gluon energies: $\omega  = 2$ GeV (left column) and $\omega = 10$ GeV (right column). The various curves in fig.~\ref{fig:AngularSpectrum-combined} are as follows: The solid (black) curve depicts the coherent spectrum off the quark, defined (neglecting for the moment the additional phase space factors) as
\beq
\label{eq:spectrum-coherent}
\mathcal{P}^\text{med}_q = \mathcal{R}_q^\text{med} - \mathcal{J}_q^\text{med} \,,
\eeq
where by the superscript ``med" we mean the sum of ``in-in", ``in-out" and ``out-out" contributions keeping in mind that we have subtracted off the vacuum (angular-ordered) coherent spectrum. For example, the total, medium-induced spectrum is given by
\beq
\label{eq:spectrum-independent}
\mathcal{R}_q^\text{med} = \mathcal{R}_q^\text{in-in} + \mathcal{R}_q^\text{in-out} \,.
\eeq
We plot this, latter contribution with a dashed (black) line in fig.~\ref{fig:AngularSpectrum-combined}. To guide the eye we have also plotted the vacuum-like spectrum at large angles (antiangular ordering \cite{MehtarTani:2010ma}) with a thick, grey line ($\propto \Theta(\theta - \theta_\qqb)/\theta$).

The numerical results for the ``dipole" regime follows indeed the expectations of ``partial decoherence." For instance, in both upper panels we observe an almost perfect cancellation of the independent component. In the soft gluon sector, upper, left panel of fig.~\ref{fig:AngularSpectrum-combined}, the coherent spectrum follows a vacuum-like distribution up to very high angles. For higher energies, upper, right panel of fig.~\ref{fig:AngularSpectrum-combined}, the independent distribution is peaked around smaller angles and the coherent spectrum is more complicated in this case. We observe that the coherent spectrum (black curve) is oscillating around the vacuum-like distribution (grey curve). Most importantly, it is also peaking around the same angles as the independent spectrum which supports the findings in sec.~\ref{sec:interferences-dipole} of a numerically suppressed `independent'-like component. In these panels it is difficult to identify the onset of coherence, since the characteristic cut-off in angle, given by $\sim (r_\perp \omega)^{-1}$, is located at large angles where the spectrum is anyhow small (this is related to the condition $\tform \ll L$ which translates to $\omega \ll \omega_c$). The role of this cut-off will anyhow be clarified when studying the spectrum integrated over angle, see fig.~\ref{fig:EnergySpectrum-dipole} below. 

In the other case, we observe the increasing role of the independent component in the ``decoherence" regime, lower panels in fig.~\ref{fig:AngularSpectrum-combined}. The angular cut-off in the vicinity of the maximum of the independent spectrum is more clearly discernible, see lower, right panel, although the momentum broadening smears this effect out. We study this particular aspect in further detail in appendix~\ref{sec:broadening}.

\begin{figure}
\centering
\includegraphics[width=0.8\textwidth]{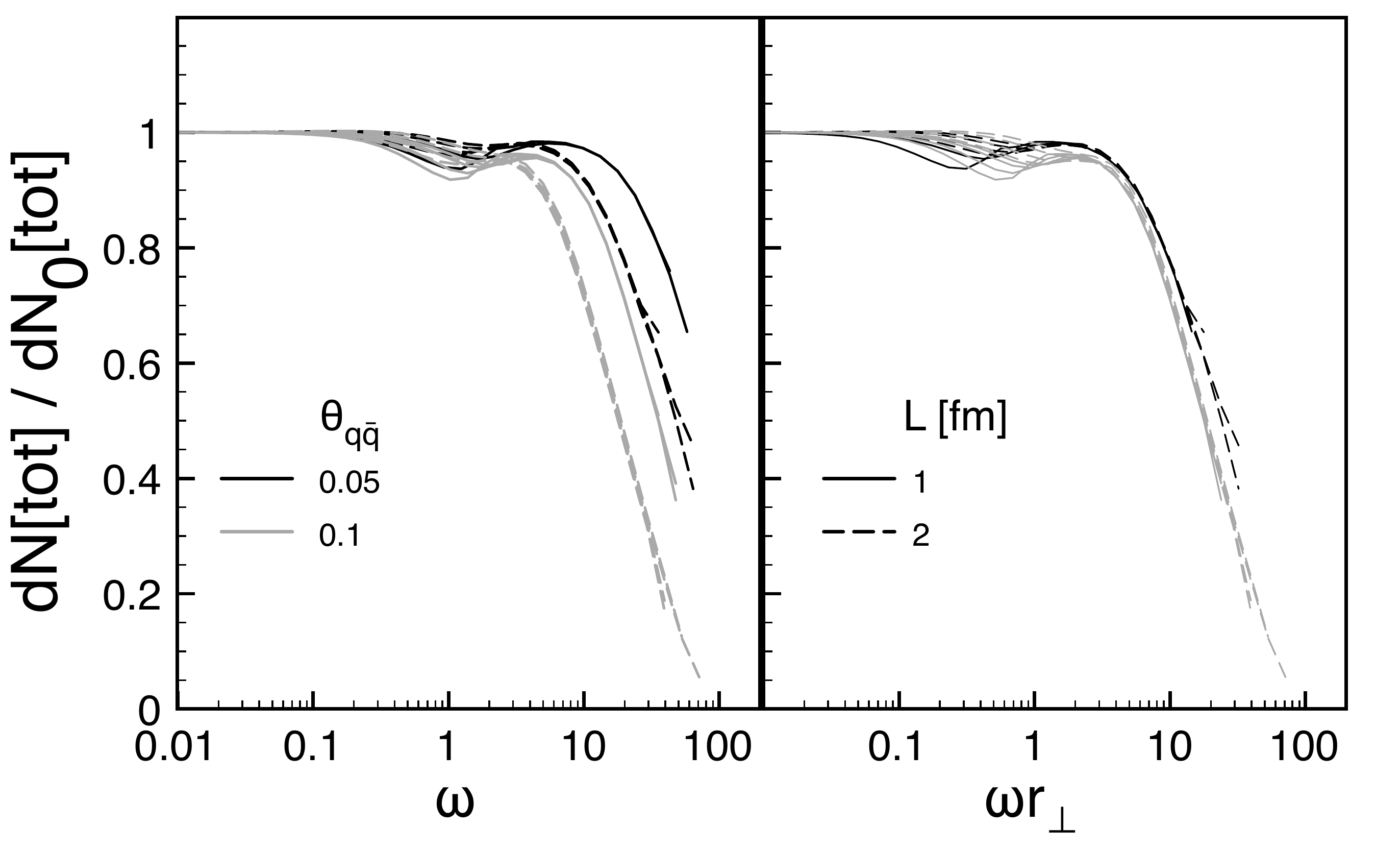}
\caption{The gluon coherent energy spectrum, $\omega dN/d\omega$, scaled by its value in the infrared limit, in the ``dipole" regime. The parameters of the calculation are denoted in the figure, additionally with $\hat q$ = 0.5, 1, 2, 3 GeV$^2$/fm (curves for different $\hat q$ at fixed $\theta_\qqb$ and $L$ overlap).}
\label{fig:EnergySpectrum-dipole}
\end{figure}
Turning now to the energy distribution, $\omega dN/d\omega$, integrated over a `physical' region of the polar angle, $\theta \in \left[0,\, \pi/2 \right]$, we aim to uncover the role of the governing hard scale of the problem $Q_\text{hard} = \max (\Qdip,\Qmed)$ via scaling of the spectrum. In fig.~\ref{fig:EnergySpectrum-dipole} (left panel) we have plotted the coherent spectrum as a function of energy, see eq.~(\ref{eq:spectrum-coherent}), for several values of $\theta_\qqb$ and $L$ such that the hard scale of the problem is $\Qdip$, see the figure for details (curves for different $\hat q$ at fixed $\theta_\qqb$ and $L$ overlap). We have scaled all the curves by their value in the soft limit \cite{MehtarTani:2011gf}. Here we again come across complex oscillatory behavior which originates from the incomplete cancellation of the independent component in the ``dipole" regime. Disregarding this detail for the moment, the general features indeed suggest a logarithmic spectrum up to a cut-off energy. After scaling the energy-variable with the hard scale in the ``dipole" regime, right panel in fig.~\ref{fig:EnergySpectrum-dipole}, we obtain an almost perfect scaling. Note also that the two sets of curves for different opening angles break apart above $\Qdip$. This is due to the presence of a maximal energy, $\omega_\text{max} = (r_\perp \theta_\qqb)^{-1}$, that arises due to the phase space restriction on the angular spectrum \cite{MehtarTani:2011gf} --- at this energy the hard scale in medium becomes smaller than the vacuum scale $|\delta \k|$ and the spectrum is fully contained within the cone delimited by the $\qqb$-pair. The study of this region is limited by numerical precision.

\begin{figure}
\centering
\includegraphics[width=0.8\textwidth]{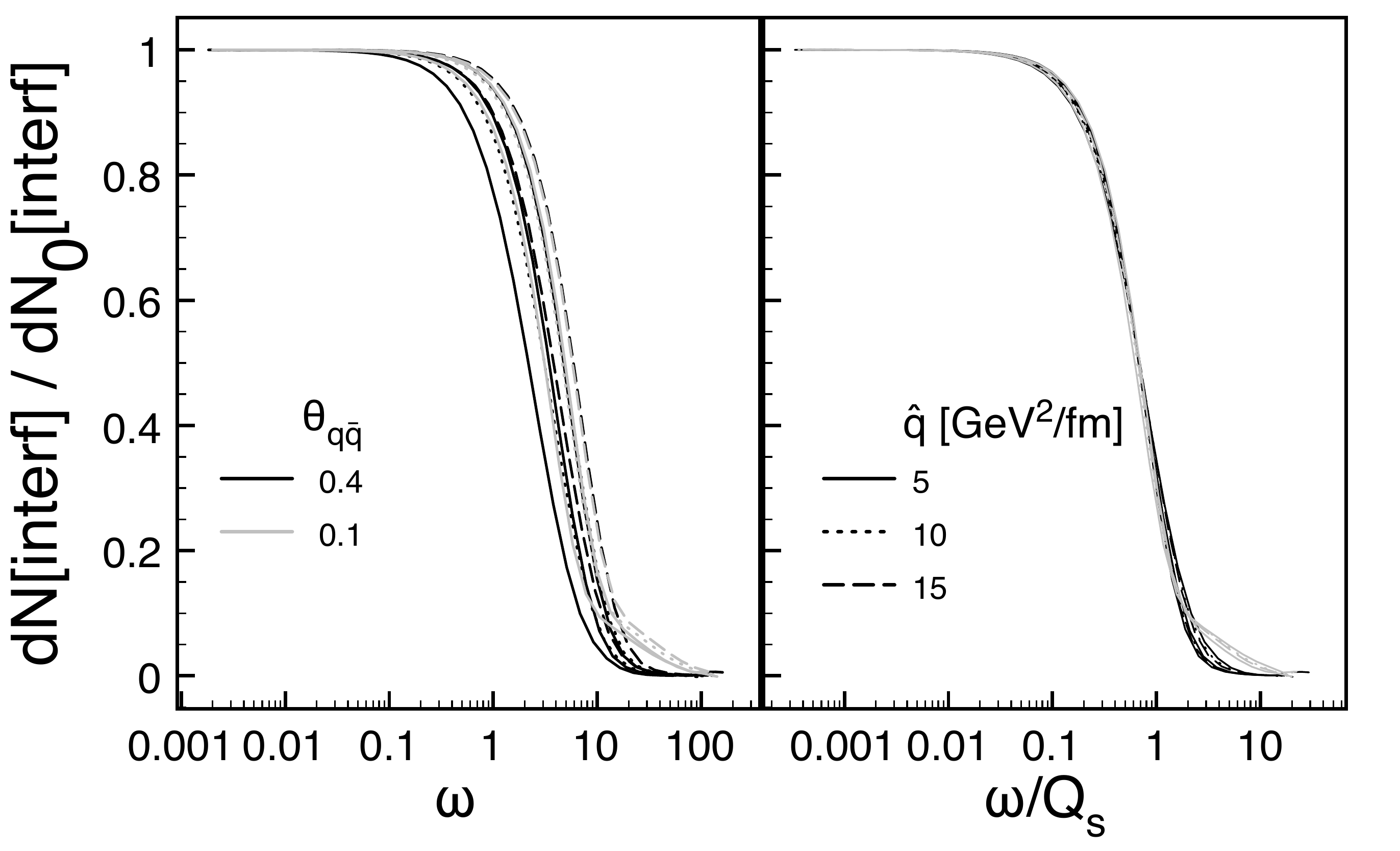}
\caption{The medium-induced interference component of the gluon energy spectrum, scaled by its value in the infrared limit, in the ``decoherence" regime. The curves are calculated with $L$ = 0.5, 1, 2 and 5 fm (curves with different $L$ are not indicated).}
\label{fig:EnergySpectrum-decoh}
\end{figure}
Clearly, from the discussion in sec.~\ref{sec:interferences-dense}, the gluon spectrum in the ``decoherence" regime is significantly less intricate since most of the components cancel and the leading behavior is quite straightforward. It is also well-known that the maximum of the energy distribution of the independent component is peaked around $\Qmed$, see, e.g., the left panel of fig.~\ref{fig:EnergySpectrum-R-scaling} in appendix~\ref{sec:broadening}. What remains to show is how the interferences behave with energy. We plot the medium-induced interference contribution, scaled its value in the soft limit \cite{MehtarTani:2011gf}, in fig.~\ref{fig:EnergySpectrum-decoh}. It displays a logarithmic behavior with energy, which originates from the $\mathcal{J}^\text{out-out}_{(1)}$ component (recall that we have subtracted the vacuum contribution), which reads
\beq
\left. \mathcal{J}^\text{out-out}_{(1)}\right|_{k_\perp < \Qmed} = - \Delta_\text{med}(L) 4\omega^2\frac{\bkappa\cdot \bbkappa}{\bkappa^2 \bbkappa^2} \,,
\eeq
for small transverse momenta. The scaling of the curves with the hard scale of the medium is apparent from the right panel of fig.~\ref{fig:EnergySpectrum-decoh}. This demonstrates that the interference contribution dominates the spectrum in the soft limit \cite{MehtarTani:2010ma,MehtarTani:2011tz}. 

In conclusion, we affirm that the numerical evaluation of the full spectrum supports the main analytical findings derived in the preceding sections. Most importantly, the role of the hardest scale of the problem has been firmly established.

\section{The octet spectrum --- reshuffling of vacuum radiation}
\label{sec:octet}

So far, we have limited our analysis to the singlet spectrum. Before concluding our investigations, let us comment briefly on the impact of our results on the general, octet spectrum. We recall that an additional component, proportional to the total color charge of the pair, is generated, see eq.~(\ref{eq:spec-general}). Generally, this contribution indicates a reshuffling of radiation ``one step backwards" in the shower and manifests the conservation of color flow. In other words, the cancellations caused by the interferences which take place for radiation off the antenna legs is reinstated as radiation coming with the color factor of the total charge --- in the octet case, of the virtual gluon prior to the $g^\ast \to \qqb$ splitting \cite{Dokshitzer:1991wu}. In pure vacuum, this component gives rise to radiation at angles larger than the opening angle of the pair and is, in turn, re-interpreted as radiation that has taken place off the initial gluon (in spite of the fact that this contribution has not explicitly been included in the calculation). This reshuffling lends support to the notion of the angular ordering of subsequent radiation. Pursuing this analogy further, we might be tempted to ask what the emerging picture for the medium-induced radiation signifies for such an idea of ordering. As a first step, this mechanism was already described in the soft limit \cite{MehtarTani:2011tz} for radiation in the presence of a medium; presently we can extend the transverse momentum range of that analysis.

Keeping in line with our previous consideration, we will also only care about radiation at angles larger than the opening angle of the $\qqb$-pair, where interferences are relevant. The contribution to the total antenna spectrum proportional to the total color charge, in our case $\propto C_A$ in eq.~(\ref{eq:spec-general}), is simply given by the interference contribution as $ \mathcal{P}_\text{total charge} \equiv \mathcal{J}$. The corresponding results for the interferences in the two different regimes are summarized in secs.~\ref{sec:interferences-dipole-summary} and \ref{sec:interferences-dense}, respectively.

In the region of relatively small momenta, $k_\perp < Q_\text{hard}$, the behavior in the two regimes is quite different. In the ``decoherence" regime the interferences are completely suppressed, a manifestation of the fact that radiation takes place independently off all color charges \cite{MehtarTani:2011tz}. In the ``dipole" regime, on the other hand, the spectrum off the total charge only consists of components that are proportional to the decoherence parameter $1-\Delta_\text{med}(L)$. Firstly, we have found a bremsstrahlung contribution, see second term in eqs.~(\ref{eq:j-in-in-dipole-3}), which is related to the decoherence of the vacuum radiation off the total color charge. This also constitutes the universal behavior in the soft limit \cite{MehtarTani:2011tz}. In addition, the `independent'-like components of $\mathcal{J}^\text{in-in}$ in the second term of eq.~(\ref{eq:j-in-in-dipole-1}) and first term in eq.~(\ref{eq:j-in-in-dipole-3}), found specifically in the ``dipole" regime, are reinstated. Note that in the limit of small opening angle, $\theta_\qqb \to 0$, we get back to the total, i.e., vacuum plus medium-induced, independent spectrum off the initial gluon.

A universal behavior, mirroring the onset of coherence at scales $k_\perp > Q_\text{hard}$ for the antenna radiation, emerges in the hard sector in the sense that the vacuum-like spectrum that is responsible for cancellations inside the antenna is reinstated, with the correct color factor, as radiation off the total charge of the pair. The $\Delta_\text{med}$ behavior of the soft sector vanishes above the hard scale set by the system, and the spectrum behaves as
\beq
\label{eq:ptotalcharge-2}
\mathcal{P}_\text{total charge} \big|_{\Qhard < k_\perp} \sim \frac{4\omega^2}{\k^2} \,,
\eeq
which is a universal behavior valid for vacuum and both regimes in the medium. Thus, vacuum radiation which is not touched by the medium is thus only allowed at angles larger than that limiting angle determined by the same hard scale. Angular ordering or, in other words, color conservation in medium is therefore manifested not simply at the opening angle of the pair but rather at $\max (Q/E, Q_\text{hard}/\omega)$, where we have written the opening angle of the pair in terms of the virtuality of the earliest gluon, $Q$, and energy of the emerging quarks, $\theta_\qqb = Q/E$.\footnote{Note that our analysis is strictly valid only for gluons that do not resolve the structure of the hard $\qqb$-pair splitting vertex, i.e., $\omega < E \theta_\qqb$.}

\section{Conclusions}
\label{sec:conclusions}

We have presented a unified picture of radiation off correlated color charges traversing a colored medium. The object of our analysis was, specifically, the single-gluon radiation off a $\qqb$-pair created in the splitting of a highly virtual gluon or photon --- limiting ourselves to soft enough radiation that can not resolve the details of the splitting process, which we therefore can treat as instantaneous. In this work we have employed the harmonic oscillator approximation to make an analytical treatment of the effects of multiple scattering with the medium feasible. This approximation breaks down for situations when atypical, hard interactions with the medium dominate, which is the case for relatively dilute media or hard gluon production. While this regime was more carefully analyzed in \cite{MehtarTani:2011gf}, the main focus of our present analysis has been on configurations when the dense, collective effects of the medium dominate. 

Including the possibility of radiation off different charges in the medium leads to a wide variety of possible configurations whose phenomenological consequences should be further investigated. Substantiating our previous findings \cite{MehtarTani:2011gf}, we have reduced this problem by identifying the relevant transverse scales which fully characterize the radiation. The hardest scale of the problem, found by
\beq
Q_\text{hard} = \max \big( r_\perp^{-1}, \Qmed, |\delta \k| \big) \,,
\eeq
determines concurrently the corresponding regime governing the emission, see tab.~\ref{tab:regimes}. In the analysis presented here, we have always assumed the two former, medium scales to be larger than $|\delta \k|$. In the opposite case, the radiation takes place independently off the antenna legs and at small angles, i.e., fully contained within the cone delimited by the $\qqb$-pair. Most interestingly, radiation above the hardest scale of the problem is strongly suppressed,
\beq
\mathcal{P}_q \big|_{\Qhard < k_\perp}  \simeq 0 \,,
\eeq
manifesting the restoration of coherence. As discussed in sec.~\ref{sec:octet}, keeping with the expectations from color conservation this contribution is reinstated as being radiated off the total color charge of the system. In this sense, the medium serves two purposes, namely to open the phase space for bremsstrahlung radiation and by inducing radiation off the color charges traversing it.

Although some of the presented results of our discussion are known in the literature (e.g., the relevance of the hardest scale of the problem was already detailed in \cite{MehtarTani:2011gf}), this analysis have also shed light on the interplay of time- and momentum-scales relevant for jet physics in medium and completes our previous studies. We have argued that the formation time both for the $\qqb$ splitting or for gluon production in medium embodies, in fact, the time it takes the pair to decohere. In a dense medium or, alternatively, for relatively soft gluon production, these times are usually very short compared to the total medium length and generally one can neglect all interferences. The emission spectra scale with the length of the medium and all color charges can radiate independently. Furthermore, in the wake of the quantum emission the particles acquire additional transverse momentum broadening that acts over large distances. In the dilute limit or, equivalently, for hard gluon radiation, the picture changes drastically. In particular, gluons with transverse momenta larger than the hardest scale of the problem can only be induced off the total charge of the system and is bremsstrahlung. The reason is that these emissions take place on very short timescales --- shorter than typical times-scales induced by the medium --- and therefore they practically are not resolved by it.

These insights will prove important for the description of the showering of highly virtual partons created in the initial hard processes in heavy-ion collisions. Our analysis, based purely on perturbative QCD, seem to indicate a hierarchy of scales which establishes a region of possible influence of medium physics. We plan to investigate these aspects further for fully dynamical settings, relevant for realistic cases in heavy-ion collisions at the LHC, in future works.

\section*{Acknowledgements}

CAS is  supported by the European Research Council grant HotLHC ERC-2001-StG-279579; by Ministerio de Ciencia e Innovaci\'on of Spain grants FPA2009-06867-E and Consolider-Ingenio 2010 CPAN. CAS is a Ram\'on y Cajal researcher. This work is supported in part by the Swedish Research Council (contract number 621-2010-3326).

\appendix

\section{The harmonic oscillator in coordinate space}
\label{sec:HOcoordinate}

In this section we explicit the expressions for the independent and interference components, see sec.~\ref{sec:ho-approximation}, in coordinate space. The former are already well-known in the literature \cite{Salgado:2003gb}. Following the decomposition in sec.~\ref{sec:formal-spectrum}, there are three contributions. First we consider the ``in-in" component, describing the situation when the emission of the gluon in both the amplitude and the complex conjugate takes place inside the medium, i.e., $t,t'<L$. We find
\begin{align}
\label{eq:j-in-in-coord}
\J^\text{in-in} &=\text{Re}\, \int_0^L dt'\int_0^{t'} dt \, \exp\left[-\frac{1}{12}\,\hat q \,\delta\n^2\, t^3+i\frac{\omega}{2} \,\delta\,\n^2 t+iAB\,\y^2-\frac{\bar \bkappa^2_\y}{4F}\right]  \nn
 &\qquad \times\frac{-2iA^2 }{F}\left\{-2i+i\left[2A(B^2+1)\,\y + \omega B\,\delta \n\right]\cdot \frac{\bar\bkappa_\y}{2F}\right.\nn
 &\qquad +  \frac{AB}{F}\left(2-\frac{\bar\kappa_\y^2}{2F}\right)+2AB\,\y^2 + \omega\,\y\cdot\delta\n \Bigg\}+\text{sym}.,
\end{align}
where $\y=\delta\n\,t$, $\bar\bkappa_\y=\bar\bkappa+2A\,\y$ and $A$ and $B$ are defined in eq.~(\ref{eq:ab-coefficients}). Finally,
\beq
\label{eq:d-broadening}
D = \frac{1}{4}\hat q (L-t') \,,
\eeq
controls the final-state transverse momentum broadening and $F=D-iAB$. In the limit $\delta\n=0$, eq.~(\ref{eq:j-in-in-coord}) reduces to 
\beq
\label{eq:r-in-in-coord}
\Rq^\text{in-in} = 2\text{Re}\, \int_0^L dt'\int_0^{t'} dt \, \exp\left[-\frac{\bkappa^2}{4F}\right] \frac{4iA^2}{F^2}\left(iD+\frac{AB}{4F}\bkappa^2\right) \,,
\eeq
and similarly for $\Rqb$ with $\bkappa \to \bbkappa$. Let us turn now to the ``in-out" contribution that describes the situation when the gluon emission takes place inside the medium in the amplitude but outside of it in the complex conjugate, or vice versa. This piece reads
\begin{align}
\label{eq:j-in-out-coord}
\J^\text{in-out}& = \text{Re}\, \int_0^L dt \, \exp\left[-\frac{1}{12}\,\hat q \,\delta\n^2\, t^3+i\frac{\omega}{2} \,\delta\,\n^2 t+i\tilde A \tilde B\,\y^2-i\frac{\bar \bkappa^2_\y}{4 \tilde A \tilde B}\right]\nn
 & \qquad \times\frac{-2\omega}{\tilde B\bar\bkappa^2} \;\bar\bkappa\cdot \left[-i2 \tilde A \tilde B\,\y+i\frac{\bar\bkappa_\y}{\tilde B} \,-i\omega\delta \n\right]+\text{sym} \,. \,,
\end{align}
where $\tilde A$ and $\tilde B$ are defined in eq.~(\ref{eq:tildeab-coefficients}). Once more, in the limit $\delta\n=0$ we find the independent component
\beq
\label{eq:r-in-out-coord}
\Rq^\text{in-out} = 2\text{Re}\, \int_0^L dt \,\frac{-i2\omega}{\tilde B^2}\exp\left[-i\frac{\bar\bkappa^2_\y}{4 \tilde A \tilde B}\right] \,.
\eeq
The ``out-out" components are already given above, see eq.~(\ref{eq:j-outout-1}) for the interferences and eq.~(\ref{eq:r-out-out-1}) for the independent spectrum.

\section{Final-state broadening}
\label{sec:broadening}

Many of the finer features of the spectrum are washed out by the final-state broadening, described by the probability distribution in eq.~(\ref{eq:p-broadening}). This is because, in contrast with the emission process which itself takes place over relatively short time-scales, the broadening can accumulate transverse momentum along the whole length of the medium and naturally involves the maximal transverse broadening, given by the so-called saturation scale of the medium $\Qmed$. It is instructive to study some of these features in more detail. We will also see that the more complex interplay of the relevant medium scales and the antenna geometry precludes a simple interpretation of one energy-independent hard scale, as sketched in sec.~\ref{sec:discussion}.

To study the medium-induced gluon spectrum in the absence of the final-state broadening, we put by hand
\beq
\mathcal{P}(\k -\k', \xi) = (2\pi)^2 \delta(\k - \k') \,,
\eeq
i.e., the particle retains the momentum it was emitted with after its formation time. In the coordinate-space representation, see appendix~\ref{sec:HOcoordinate}, one sets the parameter $D$ in eq.~(\ref{eq:d-broadening}) to zero. 

\begin{figure}
\centering
\includegraphics[width=0.8\textwidth]{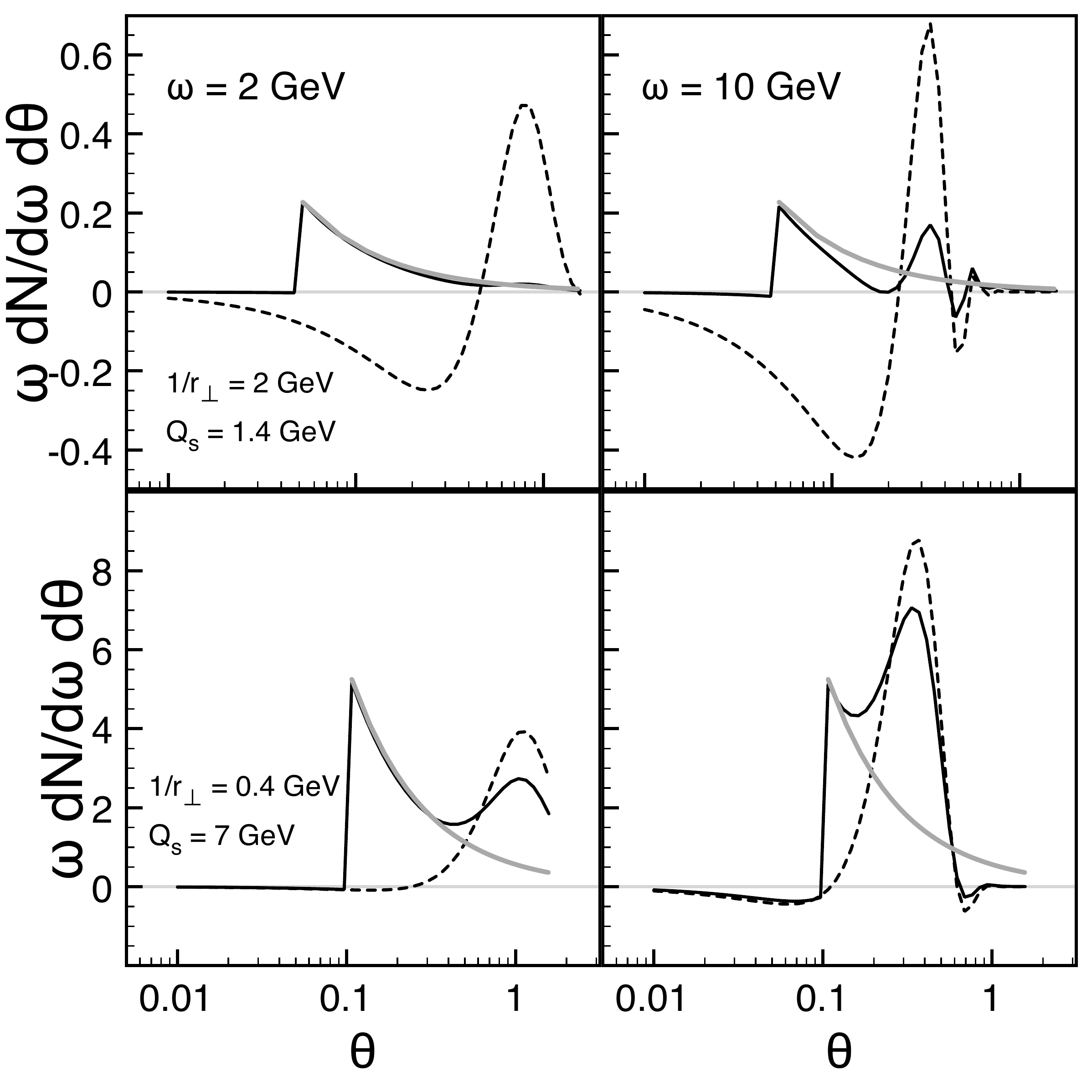}
\caption{The medium-induced gluon angular spectrum without broadening, compare with fig.~\ref{fig:AngularSpectrum-combined}. The solid (black) line corresponds to the total coherent spectrum off the quark, cf. eq.~\ref{eq:spectrum-coherent}, and the dashed (black) line is the independent spectrum, cf. eq.~(\ref{eq:spectrum-independent}). To guide the eye we have also depicted the vacuum-like spectrum at large angles with a thick, grey line ($\propto \Theta(\theta - \theta_\qqb)/\theta$). See text for further details.}
\label{fig:AngularSpectrum-combined-NoBroad}
\end{figure}
We have plotted the angular gluon spectrum without broadening in fig.~\ref{fig:AngularSpectrum-combined-NoBroad}. The curves in this figure should be compared with the corresponding curves in fig.~\ref{fig:AngularSpectrum-combined}, cf. the adjoining text for details. We note that the independent spectrum, depicted with a dashed line, is peaked more pronouncedly around its maximal value given by $\thform$, in this case, see eq.~(\ref{eq:formation-scale}) and below. In the upper, right panel of fig.~\ref{fig:AngularSpectrum-combined-NoBroad}, representing the ``dipole" regime, we observe strong oscillations of the total coherent spectrum around the peak of independent one. These features are encoded in eq.~(\ref{eq:j-in-in-dipole-3}), and discussed above. The onset of coherence at large angles is more strikingly observed in the ``decoherence" regime, see lower, right panel of fig.~\ref{fig:AngularSpectrum-combined-NoBroad}. Note that the spectrum is exactly cancelled at an angle right above $\thform$. We have checked numerically that this characteristic angle scales as expected of the hard scale $\kcoh$ relevant for this regime, see eq.~(\ref{eq:kcoh}) and the discussion leading to eq.~(\ref{eq:j-inin-dense-6}).

\begin{figure}
\centering
\includegraphics[width=0.7\textwidth]{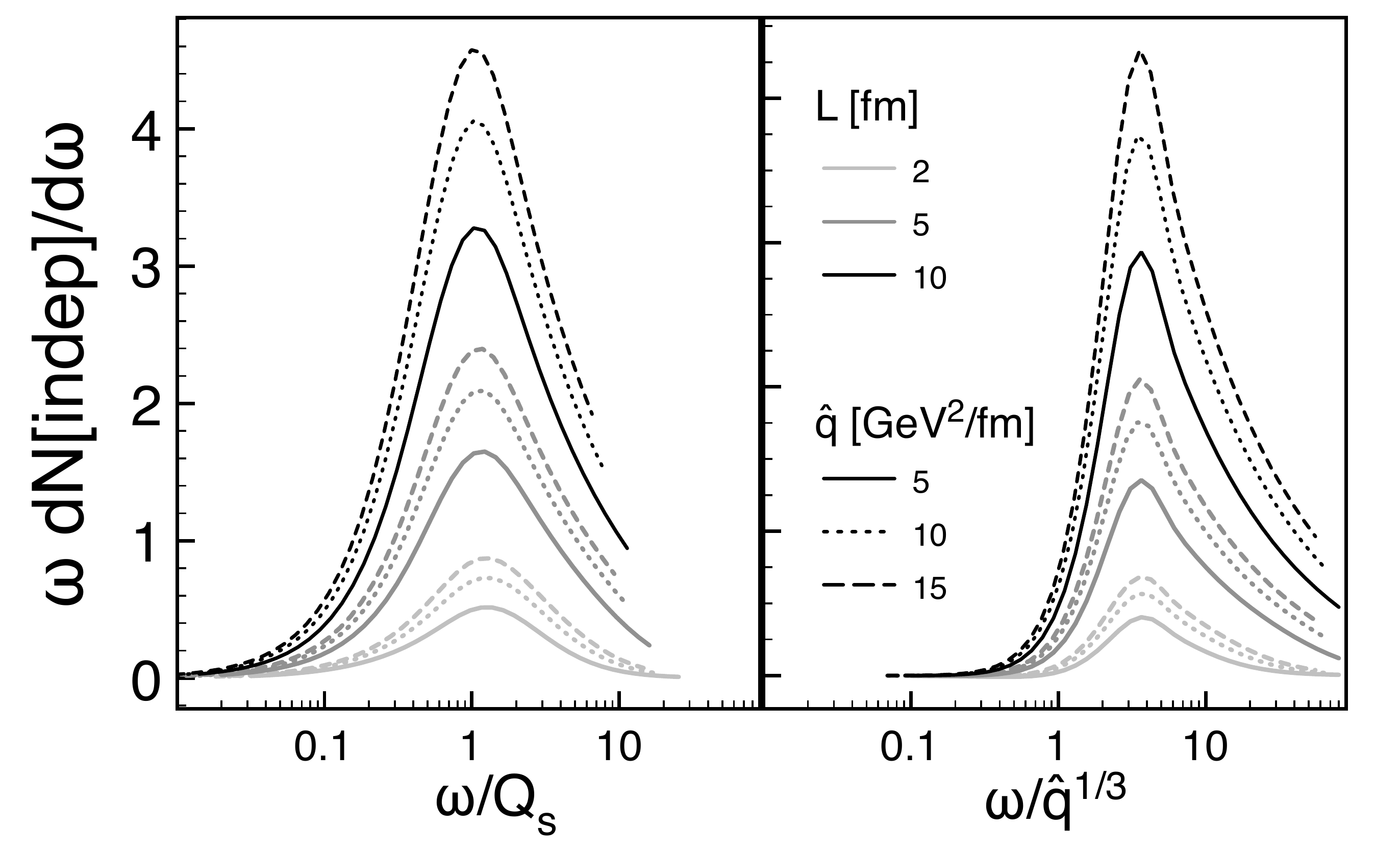}
\caption{Scaling in the independent component of the gluon energy spectrum. The values of $\hat q$ are 5, 10 and 15 GeV$^2/$fm (solid, dotted and dashed lines, respectively) and $L$ are 2, 5 and 10 fm (light-grey, dark-grey and black lines, respectively).}
\label{fig:EnergySpectrum-R-scaling}
\end{figure}
The role of the transverse momentum broadening can also be traced in the energy distribution of emitted gluons, see the left panel of fig.~\ref{fig:EnergySpectrum-R-scaling}. The peak of the energy distribution scales with the hardest medium-induced scale, namely $\Qmed$, and has a broad distribution around its central value. In the absence of broadening, the relevant scale associated with the short-distance dynamics can easily be deduced to be simply $\propto \hat q^{1/3}$. We have indeed checked that the maximum of the independent spectrum in the absence of broadening scales with this novel parameter, see the right panel of fig.~\ref{fig:EnergySpectrum-R-scaling} (see figure caption for further details). Due to the sliding hierarchy of scales in the ``decoherence" regime, see eq.~(\ref{eq:hierarchy-scales-decoherence}), no such scaling can be found for the interference contribution. 

Another, more formal, way to access the spectrum without broadening is by considering the spectrum integrated over transverse momentum. In this case, we can combine the ``in-in" and ``in-out" components of the independent gluon spectrum such that
\beq
\label{eq:r-in-in-ktint1}
\int\frac{d^2 \k}{(2\pi)^2} \mathcal{R}^\text{med}_q = -\text{Re} \frac{4i\omega}{\Omega} \int\frac{d^2 \k}{(2\pi)^2}\int_0^{\Omega L} \!dx \tan^2(x) \exp \left[ (1-i) \frac{\k^2}{2\kform^2} \tan(x) \right]  \,,
\eeq
where $\mathcal{R}^\text{med}_q = \mathcal{R}^\text{in-in}_q + \mathcal{R}^\text{in-out}_q $. Note that to be able to shift the integration limits freely in eq.~(\ref{eq:r-in-in-ktint1}) we had to assume the energy to be sufficiently large. Formally, the integration limits over the transverse momentum in eq.~(\ref{eq:r-in-in-ktint1}) are infinite. We finally arrive at the medium-induced energy spectrum, 
\beq
\label{eq:r-in-in-ktint2}
\omega \frac{dN^\text{med}}{d\omega} = \frac{2 \alpha_s C_F}{\pi}\ln\left| \cos \left( \sqrt{-i\frac{\omega_c}{\omega}} \right) \right| \,,
\eeq
where the characteristic gluon frequency is defined as $\omega_c = \hat q L^2/2$. This is the famous BDMS formula \cite{Baier:1998kq} for medium-induced soft gluon radiation. Here it is worth noting that the factor $2 C_F$ arises from the soft limit of the Altarelli-Parisi splitting function for gluon radiation off a quark, and in the general case can be replaced by $x P_{\text{q}\to \text{g}}(x)$. For an elegant generalization of eq.~(\ref{eq:r-in-in-ktint2}) to expanding media, see \cite{Arnold:2008iy}.

\bibliography{mybib}{}

\providecommand{\href}[2]{#2}\begingroup\raggedright\begin{thebibliography}{10}

\bibitem{d'Enterria:2009am}
D.~d'Enterria, {\it {Jet Quenching}},
  \href{http://xxx.lanl.gov/abs/0902.2011}{{\tt arXiv:0902.2011}}.

\bibitem{Majumder:2010qh}
A.~Majumder and M.~Van~Leeuwen, {\it {The Theory and Phenomenology of
  Perturbative QCD Based Jet Quenching}},  {\em Prog. Part. Nucl. Phys.} {\bf
  A66} (2011) 41--92, [\href{http://xxx.lanl.gov/abs/1002.2206}{{\tt
  arXiv:1002.2206}}].

\bibitem{Back:2004je}
B.~B. Back {\em et.~al.}, {\it {The Phobos Perspective on Discoveries at
  Rhic}},  {\em Nucl. Phys.} {\bf A757} (2005) 28--101,
  [\href{http://xxx.lanl.gov/abs/nucl-ex/0410022}{{\tt nucl-ex/0410022}}].

\bibitem{Arsene:2004fa}
{\bf BRAHMS} Collaboration, I.~Arsene {\em et.~al.}, {\it {Quark Gluon Plasma
  an Color Glass Condensate at Rhic? the Perspective from the Brahms
  Experiment}},  {\em Nucl. Phys.} {\bf A757} (2005) 1--27,
  [\href{http://xxx.lanl.gov/abs/nucl-ex/0410020}{{\tt nucl-ex/0410020}}].

\bibitem{Adcox:2004mh}
{\bf PHENIX} Collaboration, K.~Adcox {\em et.~al.}, {\it {Formation of Dense
  Partonic Matter in Relativistic Nucleus Nucleus Collisions at Rhic:
  Experimental Evaluation by the Phenix Collaboration}},  {\em Nucl. Phys.}
  {\bf A757} (2005) 184--283,
  [\href{http://xxx.lanl.gov/abs/nucl-ex/0410003}{{\tt nucl-ex/0410003}}].

\bibitem{Adams:2005dq}
{\bf STAR} Collaboration, J.~Adams {\em et.~al.}, {\it {Experimental and
  Theoretical Challenges in the Search for the Quark Gluon Plasma: the Star
  Collaboration's Critical Assessment of the Evidence from Rhic Collisions}},
  {\em Nucl. Phys.} {\bf A757} (2005) 102--183,
  [\href{http://xxx.lanl.gov/abs/nucl-ex/0501009}{{\tt nucl-ex/0501009}}].

\bibitem{Putschke:2008wn}
{\bf STAR} Collaboration, J.~Putschke, {\it {First Fragmentation Function
  Measurements from Full Jet Reconstruction in Heavy-Ion Collisions at
  $\sqrt{S_{_{\rm Nn}}}=200$ GeV by Star}},  {\em Eur. Phys. J.} {\bf C61}
  (2009) 629--635, [\href{http://xxx.lanl.gov/abs/0809.1419}{{\tt
  arXiv:0809.1419}}].

\bibitem{Salur:2008hs}
{\bf STAR} Collaboration, S.~Salur, {\it {First Direct Measurement of Jets in
  $\sqrt{S_{Nn}}=200$ GeV Heavy Ion Collisions by Star}},  {\em Eur. Phys. J.}
  {\bf C61} (2009) 761--767, [\href{http://xxx.lanl.gov/abs/0809.1609}{{\tt
  arXiv:0809.1609}}].

\bibitem{Lai:2009zq}
{\bf PHENIX} Collaboration, Y.-S. Lai, {\it {Probing Medium-Induced Energy Loss
  with Direct Jet Reconstruction in P+P and Cu+Cu Collisions at Phenix}},  {\em
  Nucl. Phys.} {\bf A830} (2009) 251c--254c,
  [\href{http://xxx.lanl.gov/abs/0907.4725}{{\tt arXiv:0907.4725}}].

\bibitem{Aamodt:2010jd}
{\bf ALICE} Collaboration, K.~Aamodt and C.~A. Loizides, {\it {Suppression of
  Charged Particle Production at Large Transverse Momentum in Central Pb--Pb
  Collisions at $\sqrt{S_{_{Nn}}} = 2.76$ TeV}},  {\em Phys. Lett.} {\bf B696}
  (2011) 30--39, [\href{http://xxx.lanl.gov/abs/1012.1004}{{\tt
  arXiv:1012.1004}}].

\bibitem{Aad:2010bu}
{\bf Atlas} Collaboration, G.~Aad {\em et.~al.}, {\it {Observation of a
  Centrality-Dependent Dijet Asymmetry in Lead-Lead Collisions at Sqrt(S(Nn))=
  2.76 TeV with the Atlas Detector at the Lhc}},  {\em Phys. Rev. Lett.} {\bf
  105} (2010) 252303, [\href{http://xxx.lanl.gov/abs/1011.6182}{{\tt
  arXiv:1011.6182}}].

\bibitem{Chatrchyan:2011sx}
{\bf CMS} Collaboration, S.~Chatrchyan {\em et.~al.}, {\it {Observation and
  Studies of Jet Quenching in Pbpb Collisions at Nucleon-Nucleon Center-Of-Mass
  Energy = 2.76 TeV}},  {\em Phys. Rev.} {\bf C84} (2011) 024906,
  [\href{http://xxx.lanl.gov/abs/1102.1957}{{\tt arXiv:1102.1957}}].

\bibitem{Chatrchyan:2012ni}
{\bf CMS} Collaboration, S.~Chatrchyan {\em et.~al.}, {\it {Jet Momentum
  Dependence of Jet Quenching in Pbpb Collisions at Sqrt(Snn)=2.76 TeV}},
  \href{http://xxx.lanl.gov/abs/1202.5022}{{\tt arXiv:1202.5022}}.

\bibitem{Gyulassy:1993hr}
M.~Gyulassy and X.-n. Wang, {\it {Multiple Collisions and Induced Gluon
  Bremsstrahlung in QCD}},  {\em Nucl. Phys.} {\bf B420} (1994) 583--614,
  [\href{http://xxx.lanl.gov/abs/nucl-th/9306003}{{\tt nucl-th/9306003}}].

\bibitem{Baier:1996kr}
R.~Baier, Y.~L. Dokshitzer, A.~H. Mueller, S.~Peigne, and D.~Schiff, {\it
  {Radiative Energy Loss of High Energy Quarks and Gluons in a Finite-Volume
  Quark-Gluon Plasma}},  {\em Nucl. Phys.} {\bf B483} (1997) 291--320,
  [\href{http://xxx.lanl.gov/abs/hep-ph/9607355}{{\tt hep-ph/9607355}}].

\bibitem{Baier:1996sk}
R.~Baier, Y.~L. Dokshitzer, A.~H. Mueller, S.~Peigne, and D.~Schiff, {\it
  {Radiative Energy Loss and P(T)-Broadening of High Energy Partons in
  Nuclei}},  {\em Nucl. Phys.} {\bf B484} (1997) 265--282,
  [\href{http://xxx.lanl.gov/abs/hep-ph/9608322}{{\tt hep-ph/9608322}}].

\bibitem{Baier:1998kq}
R.~Baier, Y.~L. Dokshitzer, A.~H. Mueller, and D.~Schiff, {\it {Medium-Induced
  Radiative Energy Loss: Equivalence Between the Bdmps and Zakharov
  Formalisms}},  {\em Nucl. Phys.} {\bf B531} (1998) 403--425,
  [\href{http://xxx.lanl.gov/abs/hep-ph/9804212}{{\tt hep-ph/9804212}}].

\bibitem{Zakharov:1996fv}
B.~G. Zakharov, {\it {Fully Quantum Treatment of the Landau-Pomeranchuk-Migdal
  Effect in Qed and QCD}},  {\em JETP Lett.} {\bf 63} (1996) 952--957,
  [\href{http://xxx.lanl.gov/abs/hep-ph/9607440}{{\tt hep-ph/9607440}}].

\bibitem{Zakharov:1997uu}
B.~G. Zakharov, {\it {Radiative Energy Loss of High Energy Quarks in
  Finite-Size Nuclear Matter and Quark-Gluon Plasma}},  {\em JETP Lett.} {\bf
  65} (1997) 615--620, [\href{http://xxx.lanl.gov/abs/hep-ph/9704255}{{\tt
  hep-ph/9704255}}].

\bibitem{Wiedemann:1999fq}
U.~A. Wiedemann and M.~Gyulassy, {\it {Transverse Momentum Dependence of the
  Landau-Pomeranchuk- Migdal Effect}},  {\em Nucl. Phys.} {\bf B560} (1999)
  345--382, [\href{http://xxx.lanl.gov/abs/hep-ph/9906257}{{\tt
  hep-ph/9906257}}].

\bibitem{Wiedemann:2000za}
U.~A. Wiedemann, {\it {Gluon Radiation Off Hard Quarks in a Nuclear
  Environment: Opacity Expansion}},  {\em Nucl. Phys.} {\bf B588} (2000)
  303--344, [\href{http://xxx.lanl.gov/abs/hep-ph/0005129}{{\tt
  hep-ph/0005129}}].

\bibitem{Wiedemann:2000tf}
U.~A. Wiedemann, {\it {Jet Quenching Versus Jet Enhancement: a Quantitative
  Study of the Bdmps-Z Gluon Radiation Spectrum}},  {\em Nucl. Phys.} {\bf
  A690} (2001) 731--751, [\href{http://xxx.lanl.gov/abs/hep-ph/0008241}{{\tt
  hep-ph/0008241}}].

\bibitem{Gyulassy:2000fs}
M.~Gyulassy, P.~Levai, and I.~Vitev, {\it {Non-Abelian Energy Loss at Finite
  Opacity}},  {\em Phys. Rev. Lett.} {\bf 85} (2000) 5535--5538,
  [\href{http://xxx.lanl.gov/abs/nucl-th/0005032}{{\tt nucl-th/0005032}}].

\bibitem{Gyulassy:2000er}
M.~Gyulassy, P.~Levai, and I.~Vitev, {\it {Reaction Operator Approach to
  Non-Abelian Energy Loss}},  {\em Nucl. Phys.} {\bf B594} (2001) 371--419,
  [\href{http://xxx.lanl.gov/abs/nucl-th/0006010}{{\tt nucl-th/0006010}}].

\bibitem{Arnold:2001ba}
P.~B. Arnold, G.~D. Moore, and L.~G. Yaffe, {\it {Photon Emission from
  Ultrarelativistic Plasmas}},  {\em JHEP} {\bf 11} (2001) 057,
  [\href{http://xxx.lanl.gov/abs/hep-ph/0109064}{{\tt hep-ph/0109064}}].

\bibitem{Arnold:2001ms}
P.~B. Arnold, G.~D. Moore, and L.~G. Yaffe, {\it {Photon Emission from Quark
  Gluon Plasma: Complete Leading Order Results}},  {\em JHEP} {\bf 12} (2001)
  009, [\href{http://xxx.lanl.gov/abs/hep-ph/0111107}{{\tt hep-ph/0111107}}].

\bibitem{Arnold:2002ja}
P.~B. Arnold, G.~D. Moore, and L.~G. Yaffe, {\it {Photon and Gluon Emission in
  Relativistic Plasmas}},  {\em JHEP} {\bf 06} (2002) 030,
  [\href{http://xxx.lanl.gov/abs/hep-ph/0204343}{{\tt hep-ph/0204343}}].

\bibitem{MehtarTani:2006xq}
Y.~Mehtar-Tani, {\it {Relating the Description of Gluon Production in Pa
  Collisions and Parton Energy Loss in Aa Collisions}},  {\em Phys. Rev.} {\bf
  C75} (2007) 034908, [\href{http://xxx.lanl.gov/abs/hep-ph/0606236}{{\tt
  hep-ph/0606236}}].

\bibitem{Landau:1953um}
L.~D. Landau and I.~Pomeranchuk, {\it {Limits of Applicability of the Theory of
  Bremsstrahlung Electrons and Pair Production at High-Energies}},  {\em Dokl.
  Akad. Nauk Ser. Fiz.} {\bf 92} (1953) 535--536.

\bibitem{Migdal:1956tc}
A.~B. Migdal, {\it {Bremsstrahlung and Pair Production in Condensed Media at
  High-Energies}},  {\em Phys. Rev.} {\bf 103} (1956) 1811--1820.

\bibitem{Baier:2001yt}
R.~Baier, Y.~L. Dokshitzer, A.~H. Mueller, and D.~Schiff, {\it {Quenching of
  Hadron Spectra in Media}},  {\em JHEP} {\bf 09} (2001) 033,
  [\href{http://xxx.lanl.gov/abs/hep-ph/0106347}{{\tt hep-ph/0106347}}].

\bibitem{Salgado:2003gb}
C.~A. Salgado and U.~A. Wiedemann, {\it {Calculating Quenching Weights}},  {\em
  Phys. Rev.} {\bf D68} (2003) 014008,
  [\href{http://xxx.lanl.gov/abs/hep-ph/0302184}{{\tt hep-ph/0302184}}].

\bibitem{Armesto:2009fj}
N.~Armesto, L.~Cunqueiro, and C.~A. Salgado, {\it {Q-PYTHIA: A Medium-modified
  implementation of final state radiation}},  {\em Eur.Phys.J.} {\bf C63}
  (2009) 679--690, [\href{http://xxx.lanl.gov/abs/0907.1014}{{\tt
  arXiv:0907.1014}}].

\bibitem{Lokhtin:2008xi}
I.~Lokhtin, L.~Malinina, S.~Petrushanko, A.~Snigirev, I.~Arsene, {\em et.~al.},
  {\it {Heavy ion event generator HYDJET++ (HYDrodynamics plus JETs)}},  {\em
  Comput.Phys.Commun.} {\bf 180} (2009) 779--799,
  [\href{http://xxx.lanl.gov/abs/0809.2708}{{\tt arXiv:0809.2708}}].

\bibitem{Schenke:2009gb}
B.~Schenke, C.~Gale, and S.~Jeon, {\it {Martini: an Event Generator for
  Relativistic Heavy-Ion Collisions}},  {\em Phys. Rev.} {\bf C80} (2009)
  054913, [\href{http://xxx.lanl.gov/abs/0909.2037}{{\tt arXiv:0909.2037}}].

\bibitem{Zapp:2011ya}
K.~C. Zapp, J.~Stachel, and U.~A. Wiedemann, {\it {A local Monte Carlo
  framework for coherent QCD parton energy loss}},  {\em JHEP} {\bf 1107}
  (2011) 118, [\href{http://xxx.lanl.gov/abs/1103.6252}{{\tt
  arXiv:1103.6252}}].

\bibitem{MehtarTani:2010ma}
Y.~Mehtar-Tani, C.~A. Salgado, and K.~Tywoniuk, {\it {Antiangular Ordering of
  Gluon Radiation in QCD Media}},  {\em Phys. Rev. Lett.} {\bf 106} (2011)
  122002, [\href{http://xxx.lanl.gov/abs/1009.2965}{{\tt arXiv:1009.2965}}].

\bibitem{MehtarTani:2011tz}
Y.~Mehtar-Tani, C.~A. Salgado, and K.~Tywoniuk, {\it {Jets in QCD Media: from
  Color Coherence to Decoherence}},  {\em Phys. Lett.} {\bf B707} (2012)
  156--159, [\href{http://xxx.lanl.gov/abs/1102.4317}{{\tt arXiv:1102.4317}}].

\bibitem{MehtarTani:2011gf}
Y.~Mehtar-Tani, C.~A. Salgado, and K.~Tywoniuk, {\it {The Radiation Pattern of
  a QCD Antenna in a Dilute Medium}},  {\em JHEP} {\bf 04} (2012) 064,
  [\href{http://xxx.lanl.gov/abs/1112.5031}{{\tt arXiv:1112.5031}}].

\bibitem{CasalderreySolana:2011rz}
J.~Casalderrey-Solana and E.~Iancu, {\it {Interference Effects in
  Medium-Induced Gluon Radiation}},  {\em JHEP} {\bf 08} (2011) 015,
  [\href{http://xxx.lanl.gov/abs/1105.1760}{{\tt arXiv:1105.1760}}].

\bibitem{Armesto:2011ir}
N.~Armesto, H.~Ma, Y.~Mehtar-Tani, C.~A. Salgado, and K.~Tywoniuk, {\it
  {Coherence Effects and Broadening in Medium-Induced QCD Radiation Off a
  Massive Q ${\Bar Q}$ Antenna}},  {\em JHEP} {\bf 01} (2012) 109,
  [\href{http://xxx.lanl.gov/abs/1110.4343}{{\tt arXiv:1110.4343}}].

\bibitem{Dokshitzer:1991wu}
Y.~L. Dokshitzer, V.~A. Khoze, A.~H. Mueller, and S.~I. Troian, {\it {Basics of
  Perturbative QCD}}, . Gif-sur-Yvette, France: Ed. Frontieres (1991) 274 p.
  (Basics of).

\bibitem{MehtarTani:2011jw}
Y.~Mehtar-Tani and K.~Tywoniuk, {\it {Jet Coherence in QCD Media: the Antenna
  Radiation Spectrum}},  \href{http://xxx.lanl.gov/abs/1105.1346}{{\tt
  arXiv:1105.1346}}.

\bibitem{Baier:1998yf}
R.~Baier, Y.~L. Dokshitzer, A.~H. Mueller, and D.~Schiff, {\it {Radiative
  Energy Loss of High Energy Partons Traversing an Expanding {QCD} Plasma}},
  {\em Phys. Rev.} {\bf C58} (1998) 1706--1713,
  [\href{http://xxx.lanl.gov/abs/hep-ph/9803473}{{\tt hep-ph/9803473}}].

\bibitem{Arnold:2008iy}
P.~B. Arnold, {\it {Simple Formula for High-Energy Gluon Bremsstrahlung in a
  Finite, Expanding Medium}},  {\em Phys. Rev.} {\bf D79} (2009) 065025,
  [\href{http://xxx.lanl.gov/abs/0808.2767}{{\tt arXiv:0808.2767}}].

\bibitem{CaronHuot:2010bp}
S.~Caron-Huot and C.~Gale, {\it {Finite-size effects on the radiative energy
  loss of a fast parton in hot and dense strongly interacting matter}},  {\em
  Phys.Rev.} {\bf C82} (2010) 064902,
  [\href{http://xxx.lanl.gov/abs/1006.2379}{{\tt arXiv:1006.2379}}].

\bibitem{Zapp:2012nw}
K.~C. Zapp and U.~A. Wiedemann, {\it {Coherent Radiative Parton Energy Loss
  beyond the BDMPS-Z Limit}},  \href{http://xxx.lanl.gov/abs/1202.1192}{{\tt
  arXiv:1202.1192}}.

\end{thebibliography}\endgroup
\bibliographystyle{jhep}

\end{document}